%% file: DUnlearning.tex
\documentclass[comsoc, journal, 10pt]{IEEEtran}
\usepackage[T1]{fontenc}
\IEEEoverridecommandlockouts
\usepackage{cite}
\usepackage{amsmath,amssymb,amsfonts}
\usepackage{graphicx}
\usepackage{subfig}
\usepackage{textcomp}
\usepackage{xcolor}
\usepackage{amsmath}
\usepackage{algorithm}
\usepackage{algpseudocode}
 \usepackage{multirow} 
 \usepackage{comment}
 \usepackage{makecell}
\usepackage{hyperref} 

\begin{document}

\title{PDLRecover: Privacy-preserving Decentralized Model Recovery with Machine Unlearning
}

\author{Xiangman Li, Xiaodong Wu, Jianbing Ni, Mohamed Mahmoud, and Maazen Alsabaan
\thanks{X. Li, X. Wu, and J. Ni are with the Department of Electrical and Computer Engineering and Ingenuity Labs Research Institute, Queen's University, Kingston, Ontario, Canada K7L 3N6. Email: \{xiangman.li, xiaodong.wu, jianbing.ni\}@queensu.ca.}
\thanks{M. Mahmoud is with the Department of Electrical and Computer Engineering, Tennessee Tech. University, Cookeville, TN 38505, USA. E-mail:mmahmoud@tntech.edu.}
\thanks{M. Alsabaan is with the Department of Computer Engineering, College of Computer and Information Sciences, King Saud University, Riyadh 11451, Saudi Arabia. E-mail: malsabaan@ksu.edu.sa.}}

\maketitle

\begin{abstract}
Decentralized learning is highly susceptible to poison attacks, in which malicious participants can manipulate local updates to degradate of model performance. Existing defense methods primarily focus on detecting and filtering malicious models to prevent a certain number of malicious clients from altering local models and poison the global model. However, correcting and recovering an already compromised global model remains a significant challenge. One of direct methods is to remove the malicious clients and retrain the model with the remaining clients to restore model performance. However, retraining needs substantial computational and temporal costs and cannot guarantee consistency and privacy of the original model.

In this paper, we propose a novel method called PDLRecover that can effectively recover a poisoned global model by utilizing historical information and prevent potential local model leakage.  The key challenge is to protect the shared historical models, while enabling to estimate the parameters of the recovered model for model reconstruction and recovery by the remaining clients. By leveraging the linear property of approximate Hessian matrix computation, we achieve privacy protection of the historical information based on secret sharing, preventing local model leakage during transmission and reconstruction. PDLRecover involves clients performing preparation steps, periodic steps, and final exact training to guarantee the accuracy and robustness of the recovered model. Our recovery process performs periodic exact updates to maintain accurate local curvature information, followed by a final precise update to ensure convergence quality. We demonstrate that the recovered global model is comparable to the retrained model, but with significantly reduced computational time and cost, while enabling the protection of local model parameters against privacy leakage. Our experimental results validate the accuracy of the recovered global model and the efficiency in global model recovery.
\end{abstract}

\begin{IEEEkeywords}
Decentralized Learning, Machine Unlearning, Privacy Preservation, Poison Attack.
\end{IEEEkeywords}

\section{introduction}

Decentralized learning is an emerging distributed machine learning paradigm that enables multiple clients, such as smartphones and IoT devices, to collaboratively train a global model without sharing their raw local data \cite{beltran2023decentralized}. It allows clients to perform local model training and exchange only encrypted model parameters, thereby preserving data privacy and reducing the risk of data breaches. Meanwhile, decentralized learning significantly lowers communication overhead by eliminating the need to upload large volumes of raw data, making it well-suited for distributed and bandwidth-constrained environments. 
Decentralized learning has demonstrated practical advantages in real-world applications. For example, Google’s Gboard uses federated learning for privacy-preserving text prediction, while autonomous driving systems leverage decentralized models trained on rich, on-board sensor data to recognize roads, pedestrians, and traffic signs \cite{hard2018federated}.

However, decentralized learning still faces significant privacy and security challenges. Although it avoids centralized data collection, this paradigm relies on the exchange of intermediate model parameters or updates, which can unintentionally leak sensitive information. Two major privacy attack vectors in this context are model inversion attacks and reconstruction attacks. Model inversion attacks attempt to recover original training data by reversing the outputs of a model, while reconstruction attacks exploit intermediate parameters to reconstruct a client’s local data \cite{fredrikson2014privacy}. For example, the structural patterns embedded in shared parameters may enable partial data recovery by adversarial clients.
In addition, privacy risks persist even when raw data is not explicitly exchanged. Membership inference attacks \cite{shokri2017membership} can determine whether specific data records were part of a client’s training dataset, simply by analyzing the exposed model behavior. As a result, unprotected decentralized learning systems remain vulnerable to various forms of information leakage, compromising user privacy during collaborative model training.

Moreover, malicious clients in decentralized learning can launch poison attacks by tampering with their local training data or submitting manipulated model updates, thereby degrading the performance of the global model. These attacks fall into two main categories: data poisoning, where adversaries inject harmful or misleading data into the training set; and model poisoning, where clients send crafted updates to corrupt the aggregation process or mislead the global model’s behavior \cite{goldblum2022dataset}. While existing defenses [references] primarily aim to limit the influence of a small number of attackers and detect malicious local updates, the problem of recovering a compromised model has received comparatively little attention. A conventional solution  \cite{xu2024machine} is to retrain the entire model from scratch, which is computationally expensive and time-consuming. Furthermore, in a decentralized setting, it is often impractical to ensure that all clients remain online and available for full model retraining. This underscores the need for efficient and collaborative model recovery mechanisms that do not rely on complete retraining. However, enabling clients to cooperatively restore the global model introduces additional privacy risks, as the communication and computation required for recovery may inadvertently expose sensitive information. Therefore, it is crucial to design recovery approaches that are not only effective and scalable but also privacy-preserving.

In this paper, we propose PDLRecover, a machine unlearning based privacy-preserving decentralized model recovery framework that uses clients' historical updates to mitigate the impact of malicious clients on the global model without evading the privacy of clients. Firstly, PDLRecover leverages the Hessian-vector product (HVP) computation \cite{haink2023hessian} to approximate model updates using historical gradients and parameter differences. To enhance the fidelity of the recovered global model, PDLRecover performs periodic exact updates to maintain accurate local curvature information, followed by a final precise update to ensure convergence quality. However, the exposure of historical updates leads to the potential model leakage in model recovery, which is a severe but has not been resolved yet. To prevent potential model leakage from the historical updates, PDLRecover employs shamir secret sharing \cite{shamir1979share} to allow each client to securely encode their local gradients for model protection. However, the integration of secret sharing and the HVP is not trivial. According to the linear property of approximate Hessian matrix computation \cite{berahas2016multi}, the secret shares are aggregated in an encrypted manner and used to reconstruct global statistics without revealing individual client updates during training. In the recovery phase, each client independently computes its local approximated update direction using only its own private shares. The global direction is then reconstructed collaboratively through Lagrange interpolation. As a result, PDLRecover enables decentralized model repair in the presence of partial client dropout, while preserving client privacy and avoiding directly accessing or revealing other clients’ model parameters.

\textit {Contributions.}
The main contributions of this paper are summarized in threefold:
\begin{itemize}
    \item We propose an efficient unlearning method for decentralized learning that leverages Hessian-vector product (HVP) computation and clients' historical updates to mitigate the impact of malicious or dropout clients. PDLRecover provides a technical solution for removing the influence of poisoned updates and enables the restoration of a contaminated global model without the need for full retraining.
    \item We extend the standard L-BFGS method with secret sharing to ensure privacy preservation for clients. It enables the computation of approximate Hessian matrix computation with the input of secret shares, enabling clients reconstructing local updates for unlearning during the recovery process.
    \item We provide theoretical proofs to demonstrate the feasibility of PDLRecover in recovering the global model and preserving client privacy. Additionally, experimental simulations show that PDLRecover can effectively restore model performance to a high level while maintaining overall stability.
\end{itemize}

The remaining of this paper is organized as follows. In Section \ref{Related Work}, we introduce the related work about poison attacks and machine unlearning. Section \ref{Preliminary Knowledge} proposes an overview of preliminary knowledge and problem formalization, followed by PDLRecover background in Section \ref{background}.  In Section \ref{PDLRecoversection}, we describe our detailed PDLRecover and the corresponding algorithms. Section \ref{result} shows the experimental results, followed by the security analysis of PDLRecover in Section \ref{Security Analysis}. Finally, we conclude this paper in Section \ref{conclustion}.

\section{Related Work}\label{Related Work}
In this section, we review some designs of machine unlearning and the novel poison attacks and detection methods. 
\subsection{Machine Unlearning}
Machine unlearning was originally proposed to ``forget" specific data samples, along with their influence, from a trained model \cite{bourtoule2021machine}, in order to comply with the ``Right to be Forgotten" as mandated by the General Data Protection Regulation (GDPR) \cite{zaeem2020effect}. It offers effective and reliable solutions for detecting and eliminating training data samples in response to privacy and regulatory requirements. Machine unlearning enables the generation of a model that behaves as if it has never learned the data points or samples designated for deletion at a client’s request.
Machine unlearning techniques can be broadly categorized into two types: exacting unlearning and approximate unlearning \cite{thudi2022necessity}. Exact unlearning is, for any data point $x_{i} \notin D$, the prediction of $M(x) = M'(x)$ \cite{chen2021machine}. Retraining is one of the methods of exact unlearning. This method retrains the model after removing the required data. However, this method is expensive and time-consuming, particularly for large datasets, decentralized architecture, and complex algorithms. Approximate unlearning aims to balance the cost and the accuracy of the model in training. Given an acceptable error threshold $\epsilon$, if for all $x \in D$ the difference $|M(x) - M'(x)| < \epsilon$, the unlearning is considered accurate \cite{chen2021machine}. The approximate unlearning methods include influence function method \cite{guo2019certified}, gradient modification \cite{wu2020deltagrad}, and re-optimization method \cite{golatkar2020eternal}. These methods strike a balance between operational costs and model performance by ensuring that the precision of the unlearning model remains within an acceptable range, while substantially reducing resource consumption and processing time.

Unlearning has also been explored in federated learning, where techniques originally designed for centralized settings are adapted. For example, Su et al. \cite{su2023asynchronous} followed the design of SISA \cite{bourtoule2021machine}, and Liu et al. \cite{liu2022right} further extended these concepts to design federated unlearning architectures. Their methods resemble the algorithm in \cite{golatkar2020eternal}, which leverages Fisher information \cite{martens2020new} for gradient modification.
Other federated unlearning methods are based on knowledge distillation \cite{wu2022federated} and model calibration \cite{liu2021federaser} to enable the removal of trained samples from the global model.
However, unlearning in fully decentralized learning has not been investigated. Achieving efficient unlearning in a fully decentralized model is challenging for two main reasons: First, the recovery process is time-consuming, as each client must communicate with others to exchange model updates in the absence of a central server. Second, model updates provided by malicious or dropout clients can lead to the failure of model recovery and reconstruction.

\subsection{Poison Attacks and Defenses}
Poison attacks compromise the training process of machine learning models by deliberately manipulating input data, labels, or model updates, with the goal of degrading the performance or altering the behavior of the final model. Assume there is a clean dataset $D = \{(x_1, y_1), (x_2, y_2), \dots, (x_n, y_n)\}$ for training a prediction model, where $x_i$ and $y_i$ are the data sample and the corresponding label of the $i^{th}$ data in the training dataset, and $n$ is the size of this dataset. The trainer trains a model $G$ based on the dataset $D$ and optimizes it based on a loss function $L$. Recently, poison attacks have been extensively explored, and various poison attacks have been identified in both centralized learning and decentralized learning. Zhang et al. \cite{zhang2023data} highlighted the vulnerability of EEG signal-based risk assessment systems to data poison attacks, particularly label-flipping attacks during the training stages of machine learning models. Similarly, Ovi et al. \cite{ovi2023confident} emphasized the susceptibility of models to data poison attacks when malicious clients use tainted training data. In the realm of cybersecurity, backdoors can be activated by attackers to manipulate poisoned models, as demonstrated in the study on deep source code processing models \cite{li2022poison}.

Additionally, poison attacks in federated learning systems have been explored, with a focus on generative adversarial networks (GANs) as a means of attack \cite{zhang2019poisoning}. These cases serve as reminders of the potential dangers associated with poison attacks in various contexts. Moreover, a recent study \cite{shan2024nightshade} delves into more specific types of poison attacks, such as prompt-specific poison attacks that aim to make poisoned samples visually identical to benign images with matching text prompts. This level of detail underscores the evolving sophistication of poison attack techniques and the difficulty of their identification and prevention.

To counter these evolving poison attacks, defense methods typically focus on distinguishing between benign and malicious samples. This is generally achieved by using model properties or activation statistics to determine if the model, training data, or test samples have been compromised. Various statistical methods have been applied to spot anomalies in poisoned datasets \cite{hayase2021spectre, tang2021demon}. For example, Hayase et al. \cite{hayase2021spectre} introduced a defense mechanism against backdoor attacks in machine learning models by leveraging robust statistics to detect and mitigate the impact of poisoned training samples. Li et al. \cite{li2021neural} proposed a defense against DNN backdoor attacks using neural attention distillation, which fine-tunes a backdoor-affected student network with clean data from a trusted teacher network. Liu et al. \cite{liu2018fine} assessed pruning and fine-tuning strategies for defense against backdoor attacks, concluding that a combined approach, termed “fine-pruning,” effectively reduced attack success rates without harming the accuracy of clean inputs.

Erasure techniques have also proven effective, as shown in works like \cite{zhao2020bridging} and \cite{borgnia2021strong}. These techniques aim to remove the influence of poisoned data by directly identifying and erasing corrupted parameters or activations, thus restoring the model’s integrity. Additionally, Fang et al. \cite{fang2020local} investigated the vulnerability of Byzantine-robust federated learning systems to local model poison attacks, showing how attackers can significantly raise global model error rates by corrupting client devices and manipulating local parameters. They modeled these attacks as optimization problems and evaluated them against various Byzantine-robust methods, highlighting severe potential performance issues. Li et al. \cite{li2021byzantine} integrated detection techniques with blockchain technology, using an anti-Byzantine consensus mechanism called Proof of Accuracy to validate models and ensure the learning process's integrity.

However, these methods primarily focus on identifying malicious samples or differentiating between poisoned and clean models, without considering the potential to recover compromised models. Cao et al. \cite{cao2023fedrecover} introduced the first model recovery technique that enables the server in federated learning to recover the global model after a poison attack by leveraging stored global models and clients’ model updates from each training round. The server estimates each client’s model update using historical data and restores the global model based on these estimations during the recovery process. Subsequently, Jiang et al. \cite{jiang2024towards} enhanced this approach by improving recovery speed and reducing memory usage with an efficient recovery method that employs selective information storage and adaptive model rollback. However, both methods present privacy concerns as the server must store clients’ local models for recovery, thereby exposing all historical data to the curious server, including clients’ model updates. Additionally, current designs such as \cite{cao2023fedrecover} and \cite{jiang2024towards} are tailored for federated learning, leaving fully decentralized learning unexplored.

\section{Preliminary Knowledge}\label{Preliminary Knowledge}
\subsection{Decentralized Learning}\label{Decentralized Federated Learning}
In decentralized learning \cite{roy2019braintorrent}, suppose that $n$ clients have their local dataset $D_{i}$, aiming to train a global model. Their target is to minimize the loss function $min~L(D;\textbf{w})$, where $D$ is the joint dataset of $n$ clients, i.e., $D = \sum D_{i}$, $\textbf{w}$ is the parameter of the global model, and $L$ is the loss function, such as mean square error and cross-entropy loss. Each client trains its local machine learning model using its own dataset and receives model updates from other clients to maintain the global model. The global model is updated iteratively, in each iteration consisting of three main steps:

\begin{itemize}
\item Model Initialization: Clients establish bidirectional communication with their neighboring peers by broadcasting their availability. All clients start with the same model parameters $\mathbf{w}_0$.

\item Model Training: Each client trains its local model using its own dataset and computes the local update, $g_t^i$, based on an optimization algorithm, such as the steepest gradient method or stochastic gradient descent, where $i$ denotes the client index. A client then updates its local model parameters, $\mathbf{w}_t^i$, using the learning rate $\gamma$, according to the formula $\mathbf{w}_{t+1}^i = \mathbf{w}_t^i - \gamma \cdot g_t^i$. Afterward, clients broadcast their local model parameters through secret sharing to their neighbours and receive model parameters from them.

\item Model Aggregation: Each client collects the model parameters from its neighbors and applies an aggregation rule $A$, to update the global model: $\mathbf{w}_{t+1} = A(\mathbf{w}_t^1, \mathbf{w}_t^2, \dots, \mathbf{w}_t^n)$.
    
\end{itemize}

Clients conduct these three steps to continuously update their local models. During the model aggregation step, different aggregation rules offer distinct advantages. In PDLRecover, we employ one of the most popular methods, FedAvg \cite{mcmahan2017communication}, which efficiently combines local model updates from all clients to produce a robust global model.

FedAvg is a parameter aggregation process that uses weighted averaging to update the global model. After each client collects all the neighbors' local model parameters $ \mathbf{w}_i$ and local dataset sizes $ n_i $, it calculates the weighted average of the global model parameters, where the weights are the sizes of the local datasets, following the equation as:
\begin{equation}
    \mathbf{w}_{\text{new}} = \frac{\sum_{i=1}^{K} n_i \mathbf{w}_i}{\sum_{i=1}^{K} n_i},    
\end{equation}
where $ K $ is the number of participating devices, $ \mathbf{w}_{\text{new}} $ is the new global model parameters, $ n_i $ is the size of the $ i $-th device's local dataset, and $ \mathbf{w}_i $ is $ i $-th device's local model parameters.

\input{Table_Alg/Notation}

\subsection{Shamir's Secret Sharing}
In PDLRecover, we employ a secret sharing algorithm to secure each client's local update $g_{t}^{i}$, during the collection of the aggregated model parameters $\mathbf{w}_{t}$, which are essential for model recovery. Specifically, we use the Shamir's secret sharing to facilitate secure exchange of local updates.

Shamir's secret sharing \cite{shamir1979share} is a cryptographic method that divides a secret into multiple shares, which are distributed among a group of clients. The secret can only be reconstructed when a required number of shares are combined, ensuring that individual client updates remain private. This technique enables secure aggregation during the model recovery process while protecting the confidentiality of each client’s data.

The details of Shamir's secret sharing are as follows:

\subsubsection{Secret Distribution}

A client $j$ shares a secret $ s $ among $ n $ clients, such that any $ t $ of $ n $ clients can reconstruct the secret $ s $. To achieve this, the client $j$ uses a $ t-1 $ degree polynomial to generate the shares.
\begin{itemize}
    \item Select a random polynomial 
\begin{equation}
    f(x) = a_0 + a_1 x + a_2 x^2 + \cdots + a_{t-1} x^{t-1},    
\end{equation}
    where $ a_0 = s $ is the secret, and $ a_1, a_2, \ldots, a_{t-1} $ are randomly chosen coefficients.
    \item For each client $ i $, where $ i = 1, 2, \ldots, n $, the client $j$ selects a random $ x_i $ and computes $ f(x_i) $.
    \item The share given to the client $ i $ is the point $ (x_i, f(x_i)) $.
\end{itemize}

\subsubsection{Secret Reconstruction}
Any group of at least $ t $ clients can reconstruct the secret using their shares. This is done using the Lagrange interpolation.

\begin{itemize}
    \item Collect at least $ t $ shares $ (x_1, y_1), (x_2, y_2), \ldots, (x_t, y_t) $.
    \item Use Lagrange interpolation to reconstruct polynomial as
\begin{equation}
    f(x) = \sum_{i=1}^{t} y_i \prod_{1 \leq j \leq t, j \neq i} \frac{x - x_j}{x_i - x_j}.    
\end{equation}
    \item Evaluate this polynomial at $ x = 0 $ to find the secret $
    s = f(0) = \sum_{i=1}^{t} y_i \prod_{1 \leq j \leq t, j \neq i} \frac{-x_j}{x_i - x_j}.
    $
\end{itemize}

\subsection{L-BFGS Method}
The L-BFGS algorithm \cite{matthies1979solution} plays a crucial role in efficiently computing the approximate Hessian-vector product (HVP), significantly reducing the computational complexity of large-scale optimization problems by avoiding direct Hessian matrix calculations. It achieves this by leveraging two key buffers: a global-model difference buffer, $\widetilde{W} = [\Delta \mathbf{W}_{b_1}, \Delta \mathbf{W}_{b_2}, \ldots, \Delta \mathbf{W}_{b_s}]$, which tracks changes in global model parameters across iterations, and a local update difference buffer, $\widetilde{G}^i = [\Delta G^i_{b_1},\Delta G^i_{b_2}, \ldots, \Delta G^i_{b_s}]$, which stores gradient updates in each iteration, $s$ is the buffer size. An input vector, $\mathbf{v}$, is also used in the computation process. The L-BFGS algorithm then uses $\widetilde{W}_{t}$ and $\widetilde{G}^i_t$ as inputs to generate an approximate Hessian matrix $\tilde{\mathbf{H}}^i_t$ for the $i$-th client in the $t$-th iteration. This process is described as $ \mathbf{B}^i_t = \text{L-BFGS}(\widetilde{W}_{t}, \widetilde{G}^i_t)$. 

\section{BACKGROUND}\label{background}

\subsection{Notation}
We first define the notations (shown in Table~\ref{Notation} ) that are used to construct PDLRecover.
\subsection{PDLRecover Background}
\begin{figure*}[t]
    \centering
     \includegraphics[width=\textwidth]{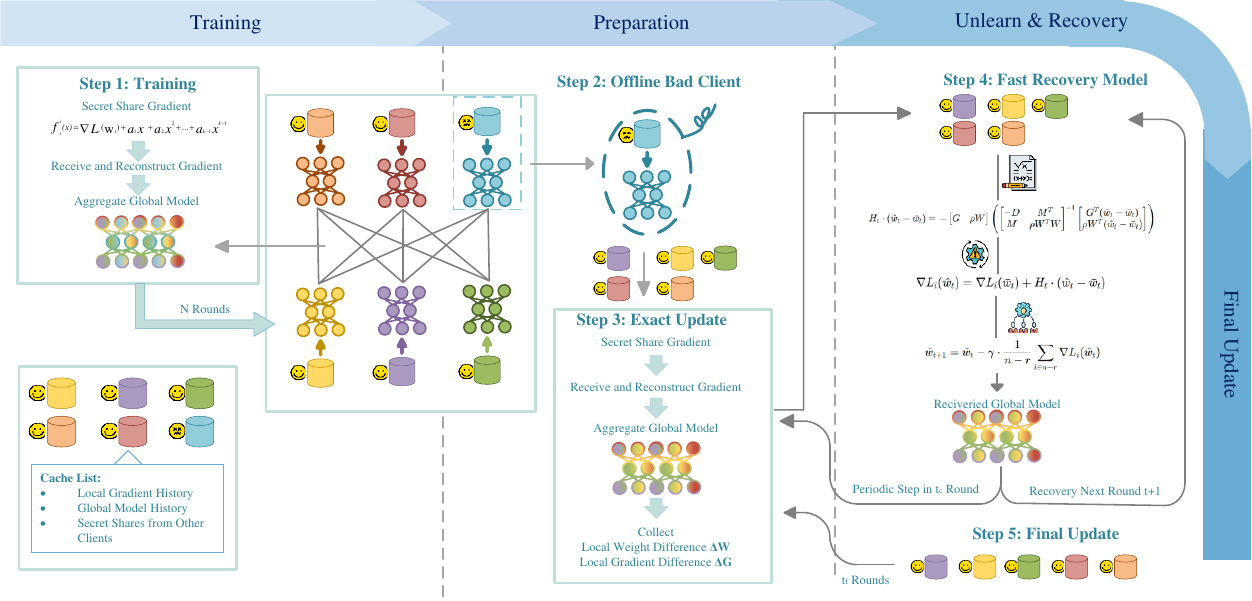}
    \caption{An overview of PDLRecover}
        \label{An overview of the PDLRecover framework}
\end{figure*}
In decentralized learning, suppose each client has a local dataset $D_{i}$ with $n$ samples to train a global machine learning model. The loss function for a client $i$ is defined as
\begin{equation}
L_{i}(\mathbf{w}) = \frac{1}{n} \sum_{j=1}^{n} L_{ij}(\mathbf{w}),
\end{equation}
where $\mathbf{w}$ is the local model parameter, and $L_{ij}(\mathbf{w})$ is the $j$-th sample's loss function for the client $i$. The gradient of $L(\mathbf{w})$ for the client $i$ is
\begin{equation}
\nabla L_{i}(\mathbf{w}) = \frac{1}{n} \sum_{j=1}^{n} \nabla L_{ij}(\mathbf{w}).
\end{equation}

Model parameters are aggregated through FedAvg in each iteration $t = 1, \ldots, T$ as
\begin{equation}
\mathbf{w}_{t+1} \leftarrow \mathbf{w}_t - \frac{\gamma_t}{n} \sum_{i \in n } \nabla L_{i}(\mathbf{w}_t),
\end{equation}
where $n$ is the number of clients and $\gamma_t$ is the learning rate at iteration $t$.

If a subset $P = \{i_{1}, i_{2}, \cdots, i_{p}\}$ of clients drop out or are identified as dishonest, submitting malicious updates, the updates from dropped clients can be removed as:
\begin{equation}
\mathbf{w}_{t+1} = \mathbf{w}_t - \frac{\gamma_t}{n-p} \left[ \sum_{i \in n } \nabla L_{i}(\mathbf{w}_t) - \sum_{i \in P } \nabla L_{i}(\mathbf{w}_t) \right],
\label{equa7}
\end{equation}
where $p$ is the size of $P$. 

However, computing $\sum_{i \in P } \nabla L_{i}(\mathbf{w}_t)$ is impossible since updates from dropped clients are unavailable. The model parameters can be updated using only the remaining clients' updates
\begin{equation}\label{Aggre}
\mathbf{w}_{t+1} = \mathbf{w}_t - \frac{\gamma_t}{n-p}  \sum_{i \notin P } \nabla L_{i}(\mathbf{w}_t).
\end{equation}

Suppose the client $i$ caches model parameters $\{\mathbf{\bar{w}}_{0},\mathbf{\bar{w}}_{1}, \cdots, \mathbf{\bar{w}}_{t} \}$ and gradients $\{\nabla L_{i}(\mathbf{w}_{0}),\nabla L_{i}(\mathbf{w}_{1}), \cdots, \nabla L_{i}(\mathbf{w}_{t})\}$ for each iteration. Using Cauchy's Mean Value Theorem, the unlearned local updated $\nabla L_{i}(\mathbf{\hat{w}}_{t}) $ for each iteration can be estimated as
\begin{equation}
\nabla L_{i}(\mathbf{\hat{w}}_{t}) = \nabla L_{i}(\mathbf{\bar{w}}_t) + \mathbf{H}_t \cdot (\mathbf{\hat{w}}_t - \mathbf{\bar{w}}_t),
\label{Cauchy mean-value theorem}
\end{equation}
where $\mathbf{H}_t$ is an integrated Hessian
\begin{equation}
\mathbf{H}_t = \int_0^1 \mathbf{H} (\mathbf{\bar{w}}_t + x (\mathbf{\hat{w}}_t - \mathbf{\bar{w}}_t)) \, dx.
\end{equation}
This approach enables accurate estimation of updates during the recovery process while maintaining efficiency and client privacy.

According to Equation~ \ref{Cauchy mean-value theorem}, we propose PDLRecover, a novel method to address the challenges posed by detected malicious or disconnected clients during distributed training. PDLRecover effectively removes compromised and dropped clients, reinitialized the global model, and restores its performance using stored historical data, thereby eliminating the need to retrain from scratch.

To facilitate efficient model recovery, each client collects valuable historical data during the training process, including the global model parameters and client model updates. PDLRecover enables each client to gather this information from neighboring clients during the training process, supporting the estimation of clients' model updates during recovery while ensuring privacy and model performance. The main framework of PDLRecover is shown in Fig. \ref{An overview of the PDLRecover framework}. 

In terms of privacy, we employ a secret-sharing mechanism to protect client data. During training, clients share their local model updates in a secret-shared manner. During the recovery phase, each client broadcasts reconstructed model updates to facilitate global model restoration. While this approach imposes certain constraints on our aggregation technique and introduces some computational overhead, the benefits of maintaining client privacy outweigh these costs.

The core of PDLRecover lies in enabling clients to communicate and collaborate in estimating global model updates during recovery in decentralized learning. Clients store not only their own training history but also global model updates, including corrupted updates introduced by malicious clients. During recovery, we estimate the model updates for each client in each iteration using Cauchy's Mean Value Theorem. Although calculating the integrals required by this theorem is complex, we overcome this challenge by estimating the Hessian matrix using the extended L-BFGS technique. While this approach necessitates certain computational and storage capabilities from clients, the benefits in terms of recovery accuracy outweigh the associated costs.

\subsection{Threat Model}\label{Threat Model}
\textbf{Attacker's Goals. }
The primary objective of an untargeted poison attack is to indiscriminately increase the global model's test error rate across a significant number of test inputs. In contrast, a targeted poison attack aims to manipulate the global model to produce inaccurate predictions specifically for certain target labels chosen by the attacker, while still maintaining accuracy for other test inputs. Backdoor attacks are a form of targeted attack that involve inserting a unique trigger, such as a specific feature pattern, into the target test inputs.

\textbf{Attacker’s Capabilities. }
An attacker has controled over specific malicious clients while maintaining the integrity of honest clients. Malicious clients can either be fabricated entities introduced by the attacker or compromised legitimate clients within the decentralized learning system. They possess the ability to send arbitrary model updates to the server, thereby impacting the overall learning process.

\textbf{Attacker’s Background Knowledge. }
The attacker's background knowledge can be classified into two distinct settings: partial-knowledge and full-knowledge settings \cite{fang2020local}.
\begin{itemize}
    \item Partial-Knowledge Setting: An attacker possesses knowledge of the global model, the loss function, and has access to local training data and model updates on the malicious clients only.
    \item Full-Knowledge Setting: 
    An attacker possesses a thorough understanding of the local training data and model updates from all clients, as well as knowledge of the aggregation rules. This level of insight significantly enhances the effectiveness of poison attacks compared to a partial-knowledge scenario, enabling the attacker to devise more potent strategies for corrupting the model.
\end{itemize}

In summary, the effectiveness of poison attacks depends on the attacker’s objectives, the degree of control over malicious clients, and the depth of knowledge about the decentralized learning system. Understanding these factors is essential for developing robust defense mechanisms. 

\subsection{Design Goals}\label{Design Goals}
Our goal is for PDLRecover to perform comparably to the drop client retraining method while significantly outperforming the historical information retraining method. Additionally, PDLRecover should minimize computational and communication overhead for clients and protect the privacy of honest participants. The key design goals include:

Accuracy: PDLRecover must recover an accurate global model from poison attacks, maintaining performance similar to the drop client retraining method. The accuracy should remain high, with minimal impact from the number of malicious clients up to a certain threshold.

Efficiency: PDLRecover should lower the computational and communication load on clients, requiring minimal rounds for clients to compute local updates while effectively adjusting their models. The server's processing and storage overhead should also be kept within practical limits.

Privacy: The privacy of clients must be safeguarded during the recovery process, even when local updates are shared for aggregation. We aim to design a privacy-preserving scheme that can be integrated with the recovery method to ensure that clients' privacy is maintained during the recovery process.

Independence from detection methods: PDLRecover should be a versatile recovery method compatible with various malicious client detection techniques. It should leverage existing models that identify client behavior to facilitate recovery and remain resilient, ensuring accuracy and stability even if some malicious clients go undetected or honest clients are mistakenly flagged.

\section{PDLRecover}\label{PDLRecoversection}
In this section, we present our PDLRecover, as shown in Algorithm~\ref{PDLRecover}, including prepareation, recovery,  periodic step, and final exact update. The proposed design integrates exact training, gradient recovery through the extended L-BFGS algorithm, and secure reconstruction of model updates through secret sharing. This hybrid strategy ensures resilience in the face of partial client dropout while preserving privacy and maintaining high recovery fidelity.

\textbf{Preparation.}
In the preparation phase, all remaining clients actively participate in federated training. At each iteration $t$, client $i$ computes the local gradient $\nabla L_i(\bar{w}_t)$ and engages in a secret sharing protocol to ensure privacy of its update. Specifically, client $i$ generates a random polynomial of degree at most $n$:
\begin{equation}
f_{i}^{t}(x) = a_{i,0}^{t} + a_{i,1}^{t}x + a_{i,2}^{t}x^2 + \cdots + a_{i,n}^{t}x^n,
\end{equation}
where the constant term encodes the private local gradient, i.e., $a_{i,0}^{t} = \nabla L_i(\bar{w}_t)$, and the remaining coefficients $a_{i,1}^{t}, a_{i,2}^{t}, \cdots, a_{i,n}^{t}$ are selected uniformly at random from a finite field $\mathbb{Z}_q$.

To construct shares, each client $j$ computes $f_i^t(x_j)$ and sends it securely to client $i$. After receiving shares from all other clients, client $i$ aggregates the sub-secret gradient values:
\begin{equation}
\nabla L^{(x_i)}(\hat{w}_t) = \sum_{j=1}^{n} f_j^t(x_i).
\end{equation}

Then, each client reconstructs the global gradient using Lagrange interpolation:
\begin{equation} \label{eq:reconstruction}
\nabla L(\bar{w}_t) = \sum_{i=1}^{n} \nabla L^{(x_i)}(\hat{w}_t) \cdot \prod_{\substack{1 \le j \le n \\ j \ne i}} \frac{0 - x_j}{x_i - x_j}.
\end{equation}

The reconstructed gradient is used to update the reference model:
\begin{equation}
\bar{w}_{t+1} = \bar{w}_t - \frac{\gamma}{n} \cdot \nabla L(\bar{w}_t).
\end{equation}

\input{Table_Alg/PDLRecover}

\input{Table_Alg/L-BFGS}

This involves calculating the precise global gradient and updating the exact model parameters, while simultaneously constructing local buffers that store  global model differences and sub-secret gradient differences, denoted by $\widetilde{W}_t$ and $\widetilde{G}_t^{(x_i)}$.  Because the recovery process spans multiple iterations, the buffers initialized during the early stages of training may become stale. Stale buffers can lead to inaccurate approximations of the Hessian matrix, erroneous model update estimates, and ultimately compromise the accuracy of the recovered global models.  To address this limitation, we require each client to repeat the preparation step every $T_r$ iterations. This regular refresh of the buffer state ensures that second-order information remains representative of the current model landscape, thereby improving both the quality and stability of the recovery process.

To support approximate second-order recovery, each client $C_j$ independently constructs local buffers. At each periodic iteration, the client stores its share of the gradient difference and the corresponding model difference:
\[
\Delta G_t^{(x_j)} = \nabla L^{(x_j)}(\hat{w}_t) - \nabla L^{(x_j)}(\bar{w}_t), \quad \Delta W_t = \hat{w}_t - \bar{w}_t,
\]
and updates:
\[
\widetilde{G}^{(x_j)}_t \gets \widetilde{G}^{(x_j)}_t \cup \{\Delta G_t^{(x_j)}\}, \quad
\widetilde{W}^{(x_j)}_t \gets \widetilde{W}^{(x_j)}_t \cup \{\Delta W_t\}.
\]

\textbf{Recovery.}
In the recovery phase, each client computes a local approximated update direction:
\begin{equation}
\hat{g}_t^{(x_j)} = \nabla L^{(x_j)}(\bar{w}_t) + \widetilde{H}^{(x_j)} (\hat{w}_t - \bar{w}_t),
\end{equation}
where $\widetilde{H}^{(x_j)}$ is computed from $\left(\widetilde{W}^{(x_j)}, \widetilde{G}^{(x_j)}\right)$ using SS-L-BFGS, an extended version of the classical L-BFGS algorithm designed to support Hessian-vector product (HVP) computation over secret shares. This ensures that the approximated update direction can be securely reconstructed from distributed shares, without revealing any client's private information.  The detailed procedure is described in Algorithm~\ref{alg:lbfgs_share}. 

Once each client has computed its private direction estimate $\hat{g}_t^{(x_j)}$, the full global direction $\hat{g}_t$ is reconstructed using Lagrange interpolation:
\begin{equation}
\hat{g}_t = \sum_{j=1}^{n} \hat{g}_t^{(x_j)} \cdot \prod_{\substack{1 \le k \le n \\ k \ne j}} \frac{0 - x_k}{x_j - x_k}.
\end{equation}

Using this aggregated update, the recovered model is updated as:
\begin{equation}
\hat{w}_{t+1} = \hat{w}_t - \frac{\gamma}{n} \cdot \hat{g}_t.
\end{equation}

During recovery, the buffers $(\widetilde{W}^{(x_j)}, \widetilde{G}^{(x_j)})$ are not modified.

\textbf{Periodic Step.}
To maintain the accuracy of the SS-L-BFGS approximation, a periodic exact step is performed every $T_r$ iterations. In this step, each client re-evaluates its gradient share at the current model state $\hat{w}_t$, computes the corresponding gradient difference $\Delta G_t^{(x_j)} = \nabla L^{(x_j)}(\hat{w}_t) - \nabla L^{(x_j)}(\bar{w}_t)$ and model difference $\Delta W_t = \hat{w}_t - \bar{w}_t$, and subsequently refreshes its local buffers by updating $\widetilde{G}^{(x_j)}_t \gets \widetilde{G}^{(x_j)}_t \cup \{ \Delta G_t^{(x_j)} \}$ and $\widetilde{W}^{(x_j)}_t \gets \widetilde{W}^{(x_j)}_t \cup \{ \Delta W_t \}$. To limit memory usage and maintain relevance, the oldest entries are discarded, preserving a fixed buffer size $s$. This periodic update guarantees that the curvature information used for Hessian approximation remains fresh and accurately reflects the evolving optimization landscape.

\textbf{Final Exact Update.}
To further stabilize the training process and eliminate accumulated approximation errors, PDLRecover concludes with $T_f$ final exact update steps. These steps mirror the standard decentralized training procedure but omit the secret sharing step. The final exact updates restore full precision to the model and ensure that it reaches a stable and accurate final state.

\section{Experiment Results}\label{result}
In this section, we first introduce the datasets, the PDLRecover implementation details, and the baseline methods we
compare with. Next, we illustrate the results that implement PDLRecover. Then, we demonstrate the performance of round number of preparation and exact training. Finally, we discuss the comparison between PDLRecover and existing methods.

\subsection{Datasets}
We use the following three datasets to implement PDLRecover.

\textbf{MNIST.} The MNIST dataset is a classic computer vision dataset widely used for image classification tasks. It contains $70,000$ grayscale images of handwritten digits, each $28x28$ pixels in size. Of these, $60,000$ images are used for training and $10,000$ for testing. Each image corresponds to a numerical label from $0$ to $9$. In this study, we used the ResNet50 model for the classification task on the MNIST dataset. To accommodate the input requirements of ResNet50, we converted the $28x28$ pixel grayscale images into $224x224$ pixel, three-channel images suitable for the model. We randomly distributed the MNIST dataset to $200$ clients for training, where the independent homogeneity is set to $0.5$, which usually ranges from $0.1$ to $1$.

\textbf{FashionMNIST.} The FashionMNIST dataset contains images of 10 different types of clothing and accessories. This dataset serves as an alternative to the MNIST dataset for more challenging image classification tasks. It consists of $70,000$ grayscale images, each $28x28$ pixels in size, with $60,000$ used for training and $10,000$ for testing. We also trained the model using $200$ clients with the same method as MNIST.

\textbf{HAR.} The HAR dataset is a standard dataset used for human activity recognition tasks. It consists of signals captured by accelerometers and gyroscopes on a smartphone and is used for recognizing $6$ different activities, including walking, walking up and down stairs, sitting, standing, and lying down. The dataset contains a total of $10,299$ samples from $30$ volunteers. We used $80\%$ of the data as the training set for the clients and the remaining $20\%$ as the test set.

\subsection{Implementation Details}
The fully decentralized learning model operates as a synchronized training and fully connected system. The clients utilize the stochastic gradient descent method to train local models, and each client employs the FedAvg algorithm to aggregate weight information for subsequent training rounds. After completing local training, clients must wait for neighboring nodes' iterations to synchronize with their own, ensuring that all clients train in unison.

For the model parameter set up, we utilize specific parameters tailored for each dataset because of the varying characteristics of different datasets. For example, MNIST and FashionMNIST are trained over $1000$ rounds with a learning rate of $1.5 \times 10^{-4}$ and a batch size of 32, whereas HAR is trained over $1000$ rounds with a learning rate of $1 \times 10^{-4}$ and a batch size of $16$.

We also set up malicious clients to build a backdoor attack. We assume the number of participating clients is $n = 200$, and the number of malicious clients is $k$, where $n = 3k - 1$. Therefore, we set the number of malicious clients to be $10$, $20$, $30$, $40$, $50$, and $60$. In MNIST and FashionMNIST, we add red stripes as the trigger to the original images, and we expect the model to recognize all image data with red stripes as birds. In HAR, we set every $20$th feature value to $0$ as the trigger value, where $0$ is the target label.

During the recovery step, the first 25 iterations are designated as the setup step, calibration is processed every 30 iterations, and the last $25$ rounds are considered as the stabilization step. We set the SS-L-BFGS buffer size to $4$.

\subsection{Baseline}
To evaluate the performance of PDLRecover, we setup two baseline methods:
\begin{itemize}
    \item Drop client retrain:
        The malicious clients are removed, the remaining clients initialized and retrain the whole model with the default model parameters and  rounds. 
    \item Historical information:
        The malicious clients are removed and the remaining clients use historical information storing in the cache to reconstruct the global. They default a same initialized model, and recovery the model directly.  
\end{itemize}


\subsection{Experiment Results}
\begin{table}
\centering
\resizebox{\linewidth}{!}
{
\begin{tabular}{|l|l|c|c|c|c|c|c|}
\hline
\multicolumn{2}{|c|}{\textbf{}} & \multicolumn{6}{c|}{\textbf{Number of clients}} \\ \hline
\textbf{Dataset Name} & \textbf{} & \textbf{10} & \textbf{20} & \textbf{30} & \textbf{40} & \textbf{50} & \textbf{60} \\ \hline
\multirow{2}{*}{MNIST} & \makecell[c]{Client \\ Retrain} & 679 & 682 & 703 & 695 & 682 & 607 \\ \cline{2-8} 
 & PDLRecover & 439 & 441 & 449 & 404 & 419 & 417 \\ \hline
\multirow{2}{*}{FashionMNIST} & \makecell[c]{Client \\ Retrain} & 707 & 706 & 707 & 711 & 708 & 710 \\ \cline{2-8} 
 & PDLRecover & 433 & 423 & 438 & 416 & 462 & 465 \\ \hline
\multirow{2}{*}{HAR} & \makecell[c]{Client \\ Retrain} & 1235 & 1212 & 1227 & 1217 & 1246 & 1250 \\ \cline{2-8} 
 & PDLRecover & 875 & 887 & 867 & 893 & 886 & 901 \\ \hline
\end{tabular}
}
\caption{Running Time (seconds) for MNIST, FashionMNIST, and HAR with attack}
\label{Running Time (seconds) for MNIST, FashionMNIST, and HAR with attack}
\vspace{-1.0em}
\end{table}

\begin{table}
\centering
\resizebox{\linewidth}{!}
{
\begin{tabular}{|l|l|c|c|c|c|c|c|}
\hline
\multicolumn{2}{|c|}{\textbf{}} & \multicolumn{6}{c|}{\textbf{Number of clients}} \\ \hline
\textbf{Dataset Name} & \textbf{} & \textbf{10} & \textbf{20} & \textbf{30} & \textbf{40} & \textbf{50} & \textbf{60} \\ \hline

\multirow{2}{*}{MNIST} & \makecell[c]{Client \\ Retrain} & 682 & 686 & 694 & 691 & 675 & 672 \\ \cline{2-8} 
 & PDLRecover & 447 & 435 & 454 & 413 & 422 & 416 \\ \hline
 
\multirow{2}{*}{FashionMNIST} & \makecell[c]{Client \\ Retrain} & 702 & 712 & 704 & 705 & 707 & 718 \\ \cline{2-8} 
 & PDLRecover & 426 & 414 & 428 & 426 & 461 & 461 \\ \hline
 
\multirow{2}{*}{HAR} & \makecell[c]{Client \\ Retrain} & 1241 & 1249 & 1223 & 1218 & 1248 & 1263 \\ \cline{2-8} 
 & PDLRecover & 871 & 879 & 876 & 897 & 876 & 902 \\ \hline
\end{tabular}
}
\caption{Running Time (seconds) for MNIST, FashionMNIST, and HAR only drop clients}
\label{Running Time (seconds) for MNIST, FashionMNIST, and HAR only drop clients}
\vspace{-2.0em}
\end{table}

\begin{figure*}[t]
    \centering
    \begin{minipage}[b]{0.32\textwidth}
        \includegraphics[width=\textwidth]{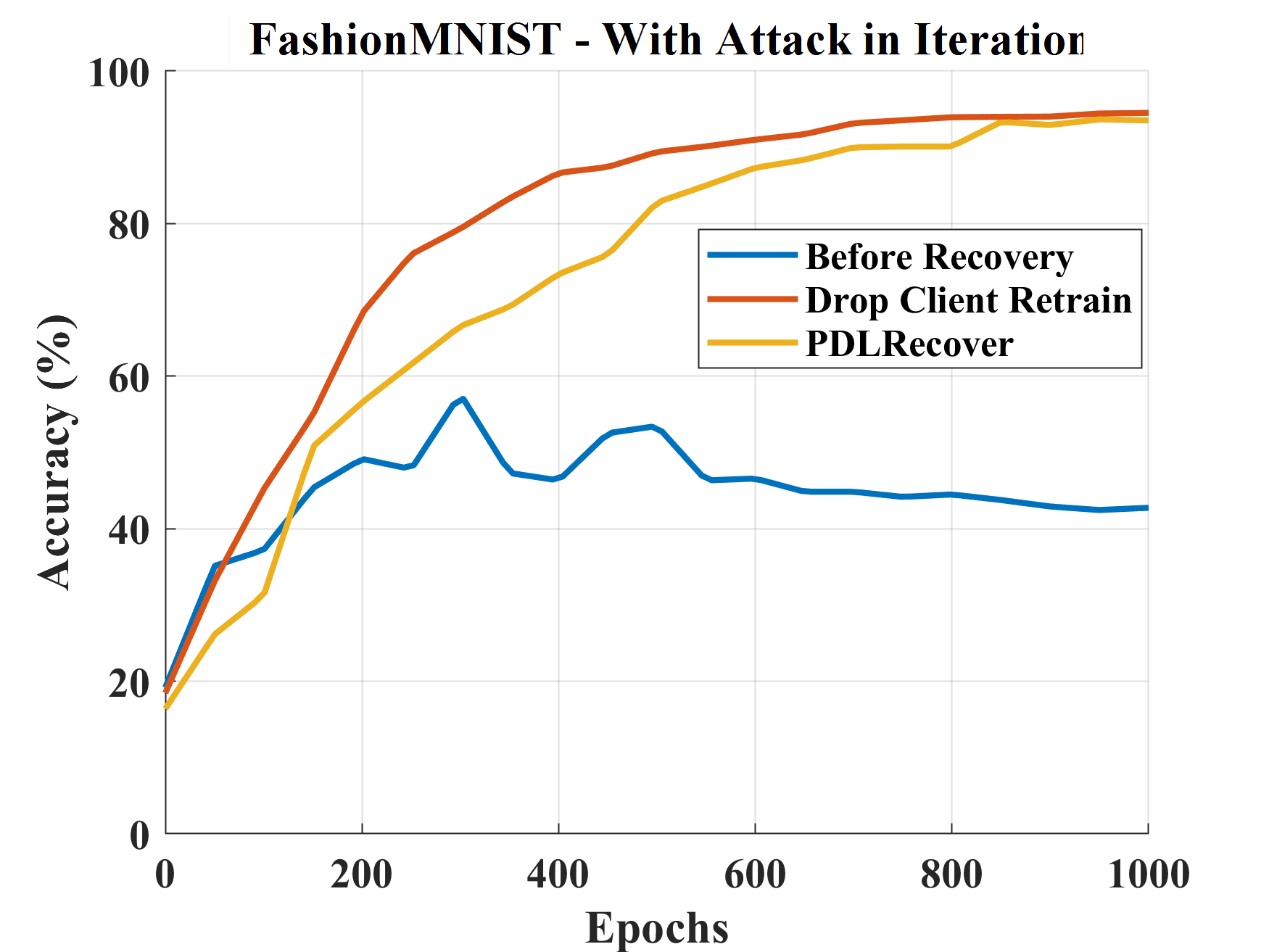}
        \label{FashionMNIST with attack in iteration}
    \end{minipage}
    \hfill
    \begin{minipage}[b]{0.32\textwidth}
        \includegraphics[width=\textwidth]{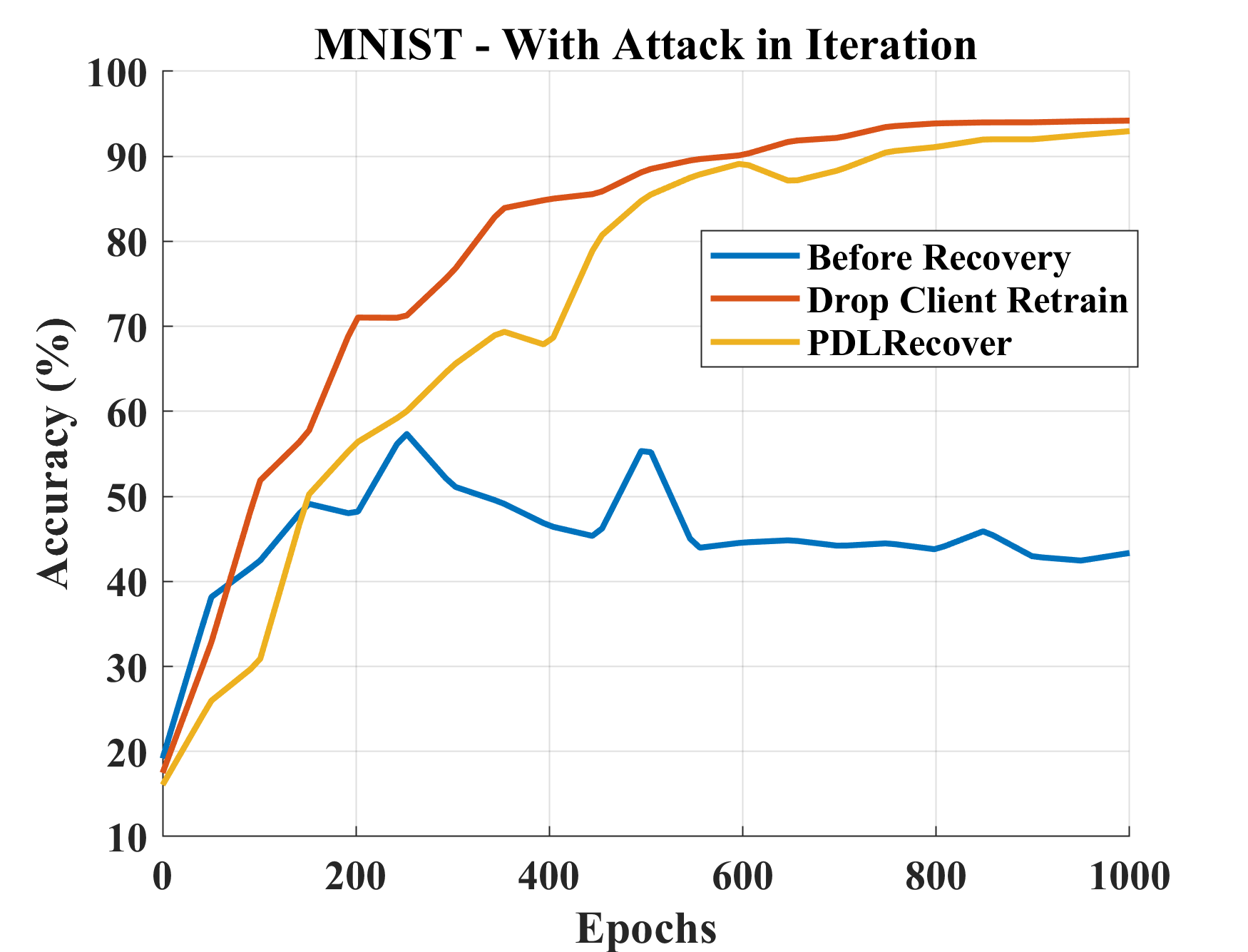}
        \label{MNIST with attack in iteration}
    \end{minipage}
    \hfill
    \begin{minipage}[b]{0.32\textwidth}
        \includegraphics[width=\textwidth]{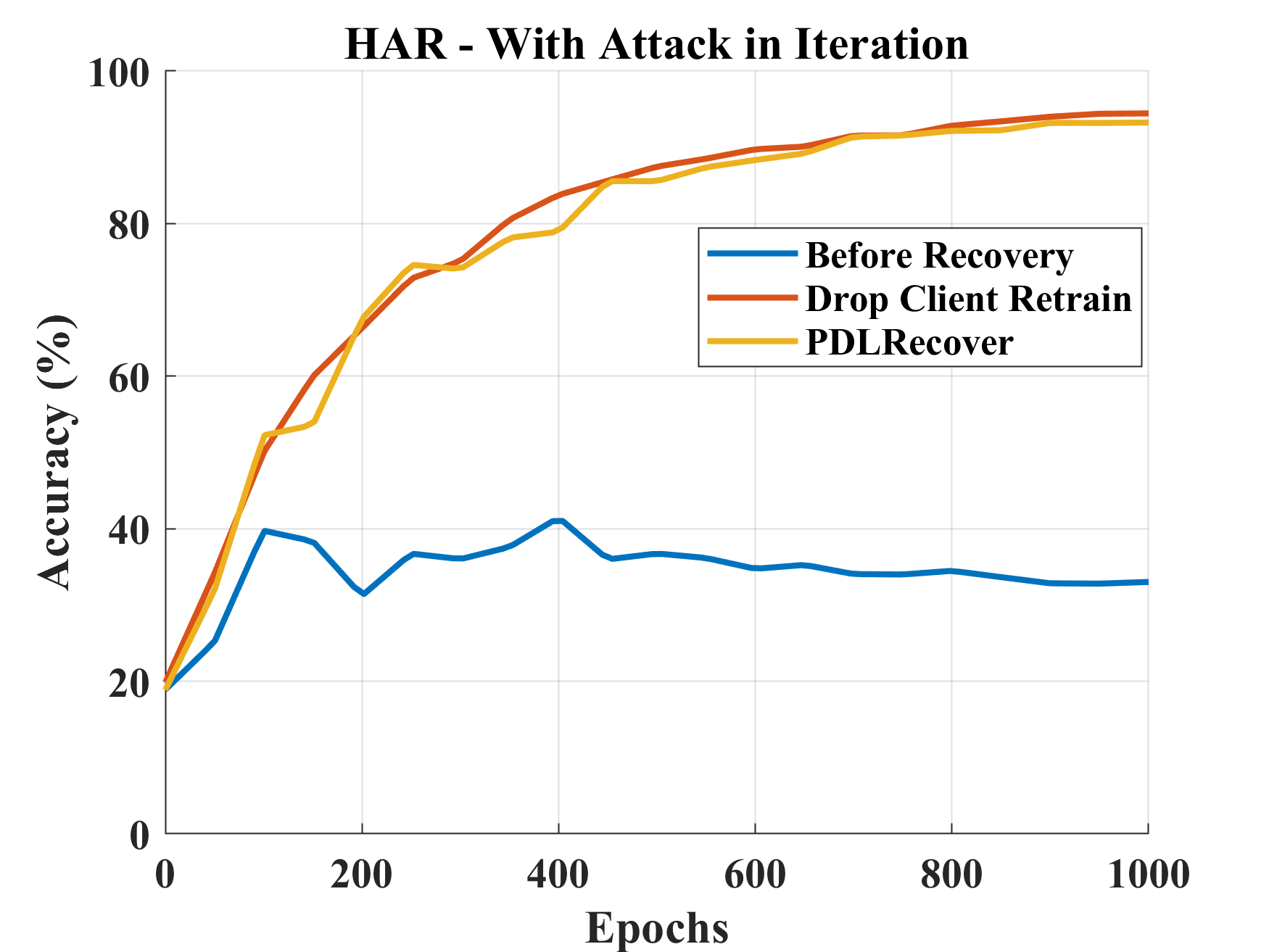}
        \label{HAR with attack in iteration}
    \end{minipage} 
    \caption{Accuracy of the recovery strategy under attack on MNIST, FashionMNIST, HAR datasets}
    \label{The accuracy of the recovery strategy under attack on MNIST, FashionMNIST, HAR datasets}
    \vspace{0.2cm}
\end{figure*}

\begin{figure*}[t]       
    \begin{minipage}[b]{0.32\textwidth}
        \includegraphics[width=\textwidth]{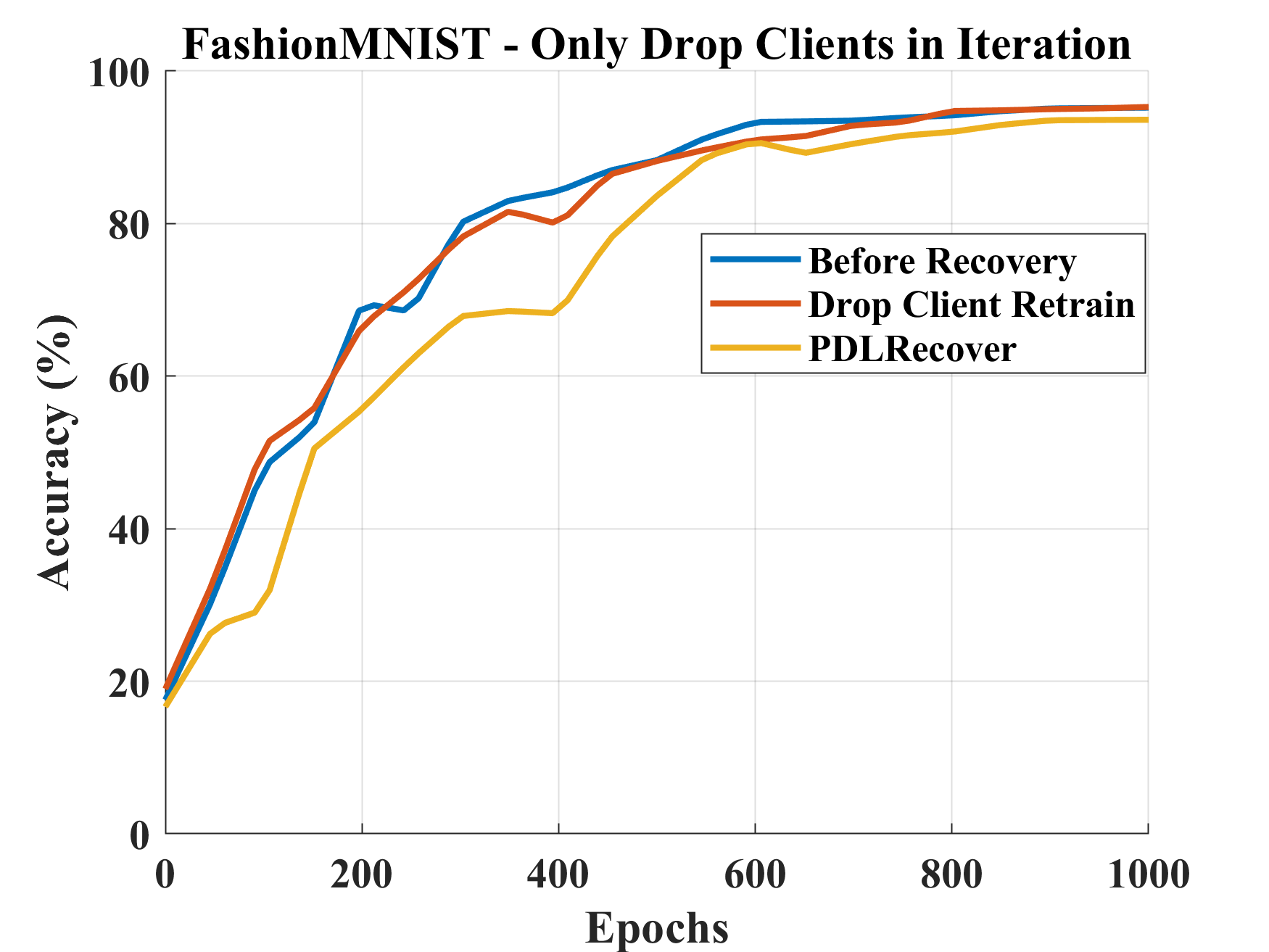}
    \end{minipage}
    \hfill
    \begin{minipage}[b]{0.32\textwidth}
        \includegraphics[width=\textwidth]{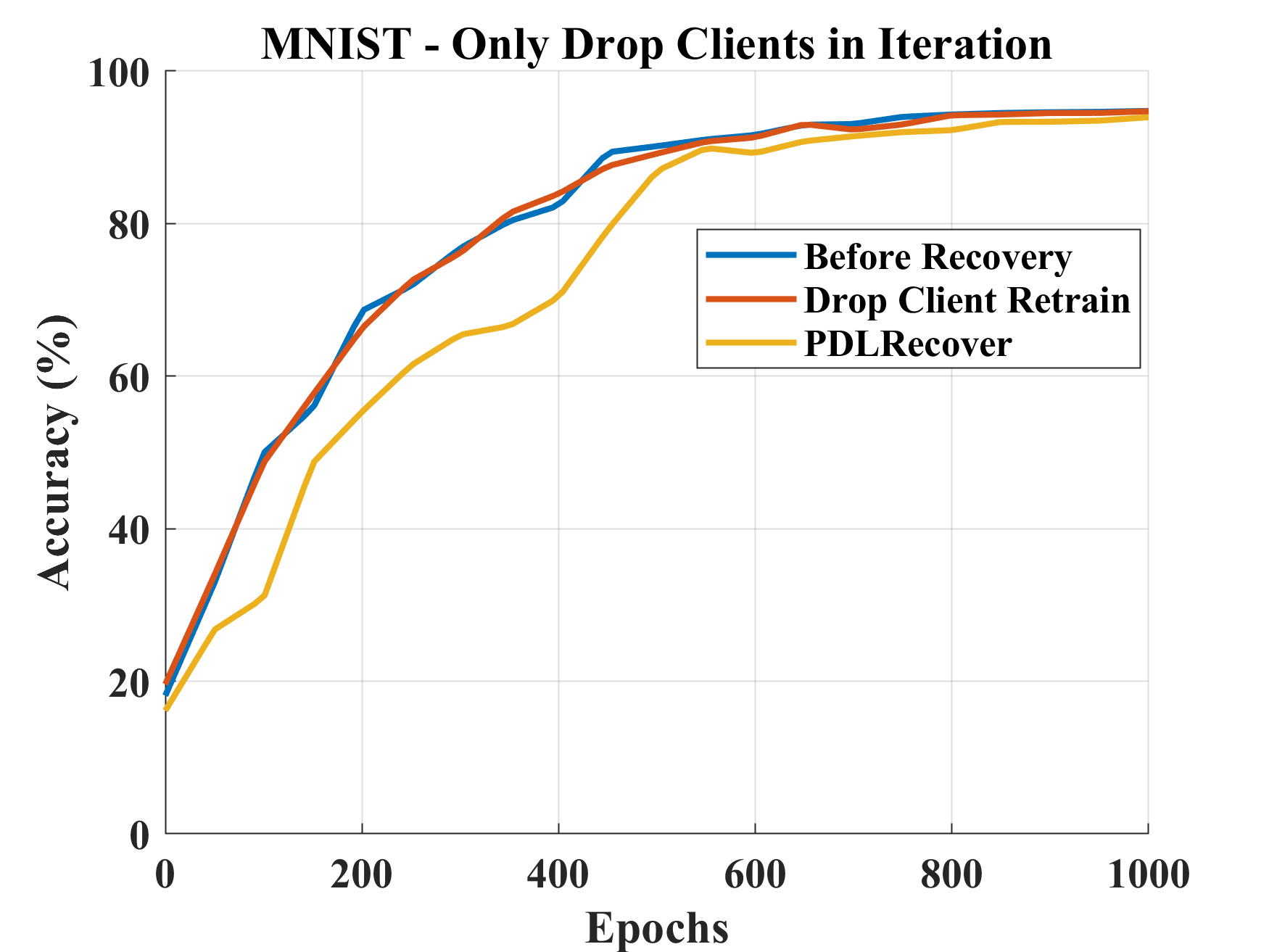}
    \end{minipage}
    \hfill
    \begin{minipage}[b]{0.32\textwidth}
        \includegraphics[width=\textwidth]{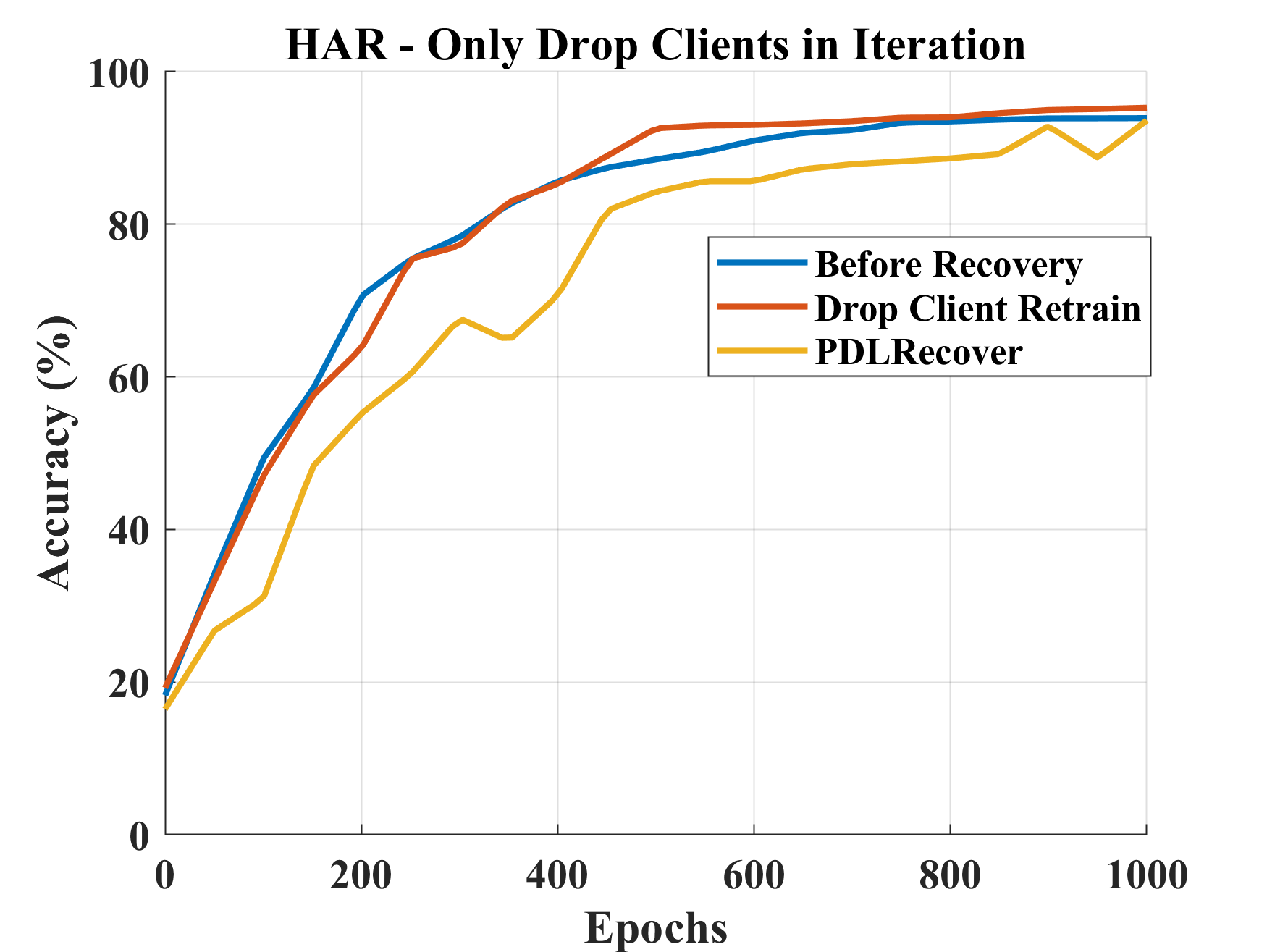}
    \end{minipage}
    \caption{Accuracy of the recovery strategy under client drop on MNIST, FashionMNIST, HAR datasets}
    \label{The accuracy of the recovery strategy under client drop on MNIST, FashionMNIST, HAR datasets}
\vspace{-1.0em}
\end{figure*}

\begin{figure*}[t]
    \centering
    \begin{minipage}[b]{0.3\textwidth}
        \includegraphics[width=\textwidth]{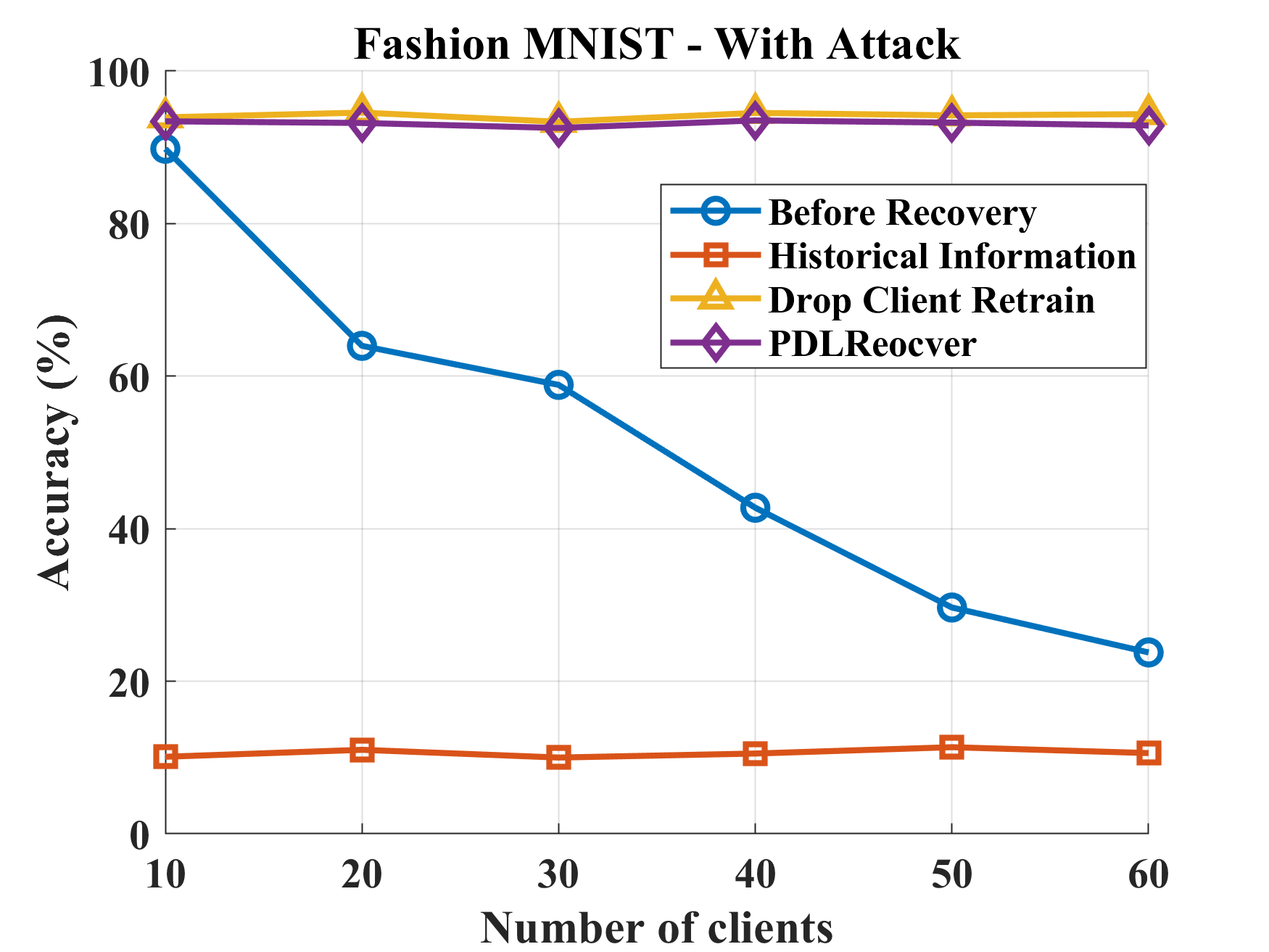}
    \end{minipage}
    \hfill
    \begin{minipage}[b]{0.3\textwidth}
        \includegraphics[width=\textwidth]{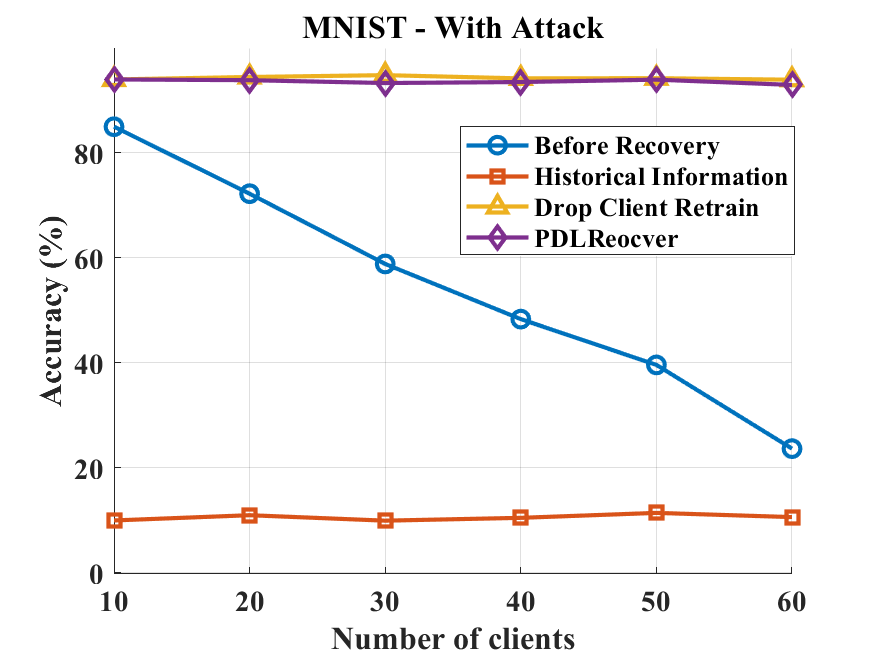}
    \end{minipage}
    \hfill
    \begin{minipage}[b]{0.3\textwidth}
        \includegraphics[width=\textwidth]{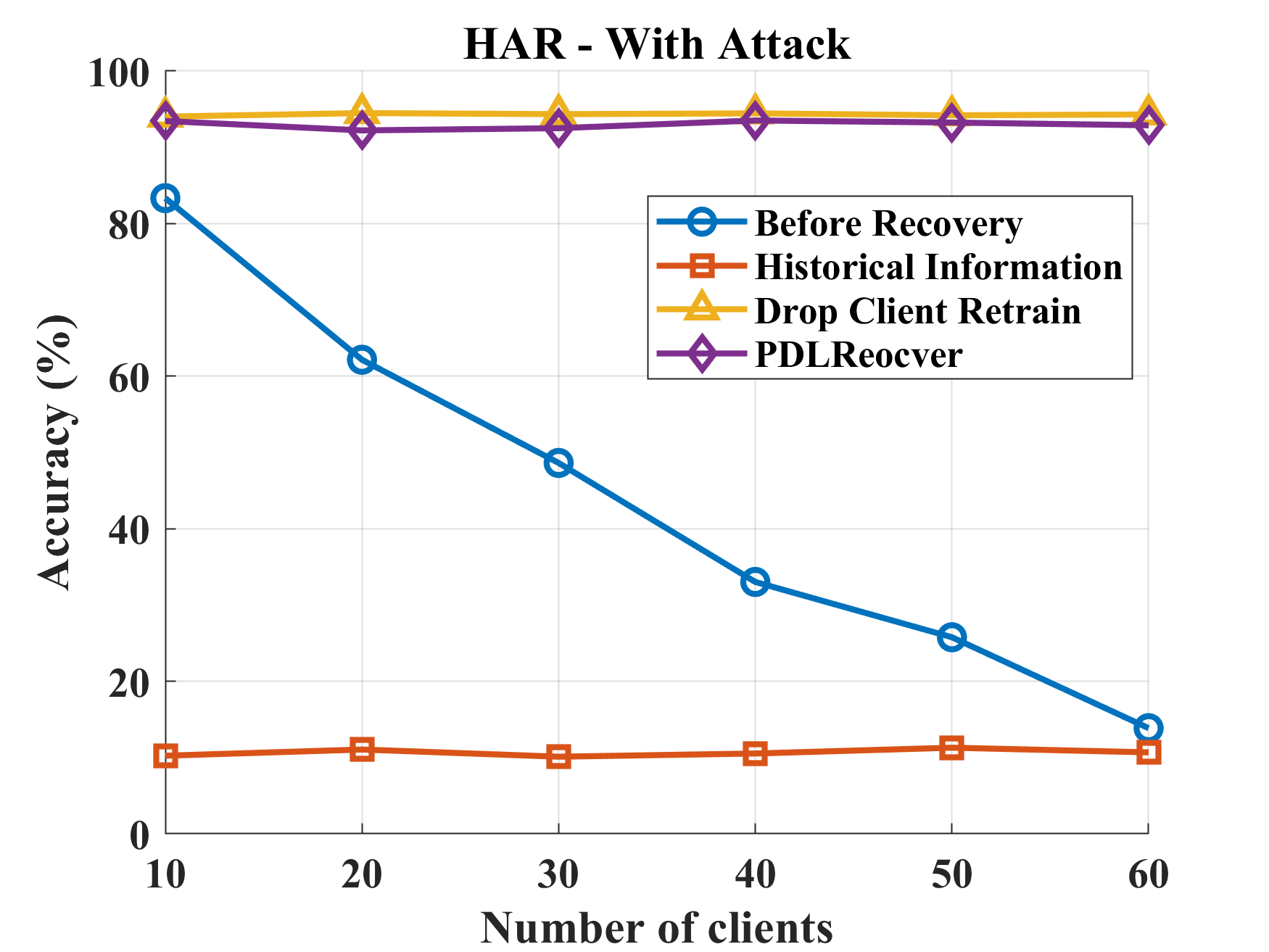}
    \end{minipage}

    \vspace{0.2cm}
    \caption{Model accuracy under different recovery strategy with different mumber of malicious clients}
    \label{Model accuracy under different recovery strategy with different malicious client number}
\end{figure*}
\begin{figure*}[t]
    \begin{minipage}[b]{0.3\textwidth}
        \includegraphics[width=\textwidth]{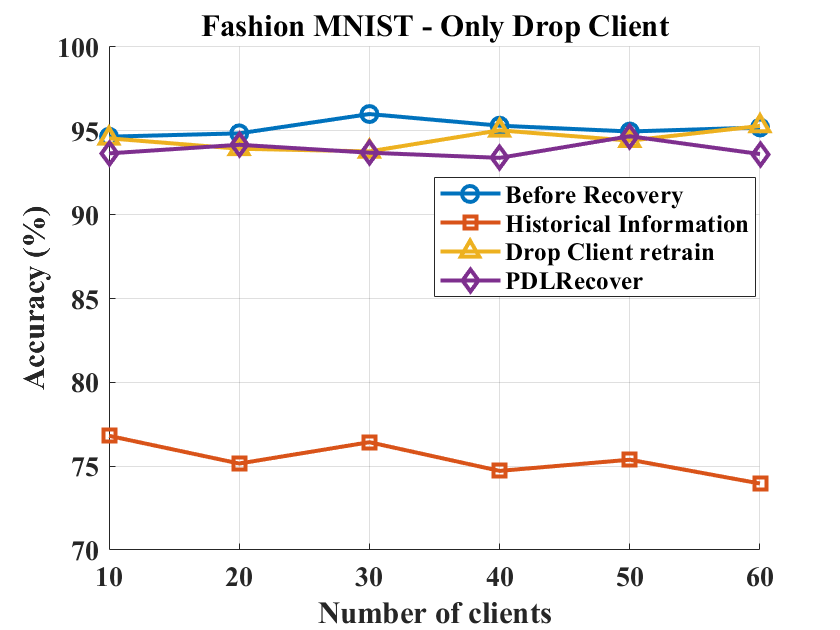}
    \end{minipage}
    \hfill
    \begin{minipage}[b]{0.3\textwidth}
        \includegraphics[width=\textwidth]{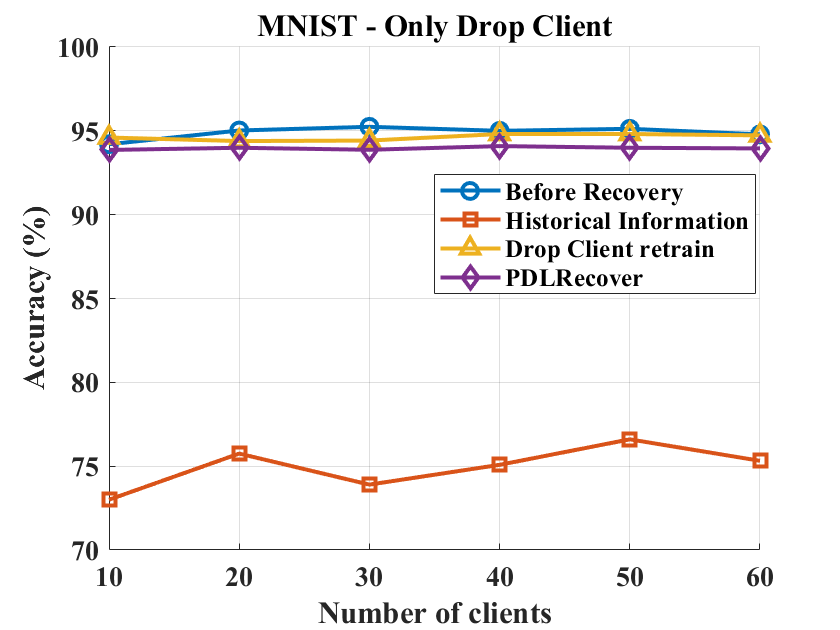}
    \end{minipage}
    \hfill
    \begin{minipage}[b]{0.3\textwidth}
        \includegraphics[width=\textwidth]{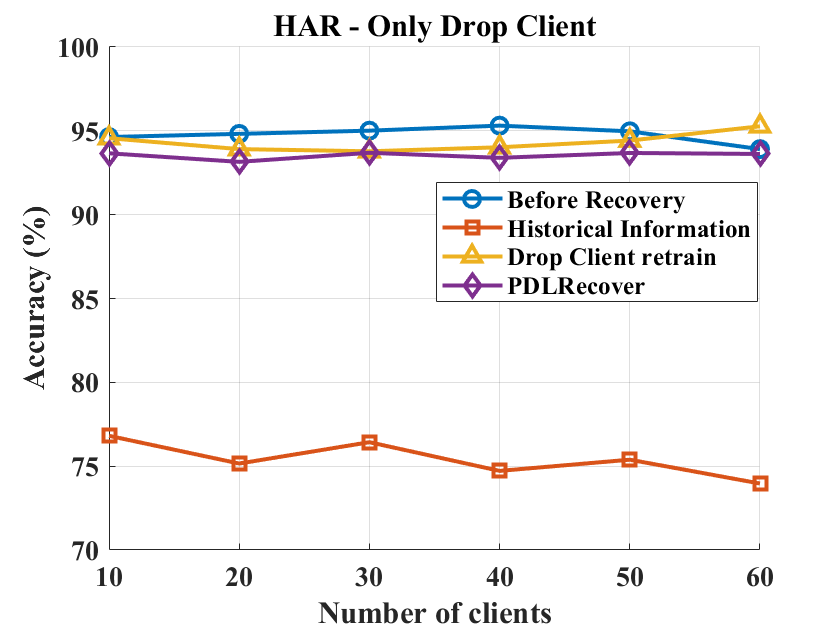}
    \end{minipage}
    \caption{Model accuracy under different recovery strategy with different drop client number}
    \label{Model accuracy under different recovery strategy with different drop client number}
\end{figure*}

Fig. \ref{The accuracy of the recovery strategy under attack on MNIST, FashionMNIST, HAR datasets} shows the accuracy of poison attacks, Drop Client Retrain, Model Before Recovery, and PDLRecover for the MNIST, FashionMNIST, and HAR datasets. It can be observed that the Drop Client Retrain and PDLRecover methods can recover the poisoned model, and PDLRecover achieves a similar effect as the Drop Client Retrain method. However, the Model Before Recovery results in lower accuracy, around $50\%$. To confirm the efficiency of PDLRecover, we also show the running time for different methods in Table~\ref{Running Time (seconds) for MNIST, FashionMNIST, and HAR with attack}, where it can be noticed that PDLRecover saves $33\%$ of the running time on average compared to the Drop Client Retrain method.

Fig. \ref{The accuracy of the recovery strategy under client drop on MNIST, FashionMNIST, HAR datasets} show the accuracy of Model Before Recovery , Drop Client Retrain, and PDLRecover for the MNIST, FashionMNIST, and HAR datasets when all clients are honest. When the dropped clients are removed and the remaining clients apply Drop Client Retrain and PDLRecover, both methods maintain better performance, with accuracy similar to the Model Before Recovery result. However, Table~\ref{Running Time (seconds) for MNIST, FashionMNIST, and HAR only drop clients} shows that the running time of PDLRecover is $35\%$ lower on average compared to the Drop Client Retrain method.

Fig. \ref{Model accuracy under different recovery strategy with different malicious client number} shows the effect of different numbers of malicious clients on the models. It can be observed that the accuracy of the poisoned model linearly decreases with the increase in the number of malicious clients. When the number of malicious clients reaches its maximum, the training accuracy of the MNIST and FashionMNIST training sets stays around $24\%$, while the HAR training set accuracy is around $13\%$, indicating that the increase in malicious clients greatly impacts the model's performance.

At the same time, the number of malicious clients does not affect the model's final performance when we recover the poisoned model using retraining and PDLRecover methods. The accuracy of the Drop Client Retrain method stays around $93\%$ in the MNIST and FashionMNIST datasets, while it remains around $94\%$ in HAR. PDLRecover stays around $91\%$ in the MNIST and FashionMNIST datasets and around $92\%$ in HAR. However, according to Table~\ref{Running Time (seconds) for MNIST, FashionMNIST, and HAR with attack}, PDLRecover can save $35\%$ of the running time.

Fig. \ref{Model accuracy under different recovery strategy with different drop client number} show all the clients are honest, maximum $60$ clients want to dropped out, the remaining clients use the methods of drop client retrain, historical information, or PDLRecover to remove the impact of the dropped clients. Our experiments show that the drop client retrain and PDLRecover can keep the performance of the model, but the historical information method decreases the accuracy of the model to around 75\%. At the same time, Table~\ref{Running Time (seconds) for MNIST, FashionMNIST, and HAR only drop clients} shows the running time of drop client retrain and PDLRecover, and PDLRecover still can save around 30\% of the running time. 
Therefore, PDLRecover can more efficiently eliminate the impact of dropped clients on the whole model and maintain the performance.
\begin{table}
\centering
\begin{tabular}{|c|c|c|c|c|c|}
\hline
\textbf{Datasets} & \textbf{Retrain} & \textbf{\cite{guo2019certified}} & \textbf{\cite{zhang2023fedrecovery}} & \textbf{\cite{cao2023fedrecover}} & \textbf{Ours} \\
\hline
MNIST & 95.4 & 95.2 & 95.3 & 94.7 & \textbf{93.9} \\
\hline
FashionMNIST & 94.9 & 86.5 & 87.4 & 85.8 & \textbf{93.2} \\
\hline
HAR & 94.6 & 84.4 & 86.5 & 87.2 & \textbf{93.4} \\
\hline
\end{tabular}
\caption{Accuracy(\%) of unlearning methods comparison}
\label{The accuracy of unlearning methods, comparison}
\end{table}

\subsection{Round of Recovery Preparation and Exact Training}
\begin{figure}[t]
  \centering
  \includegraphics[width=0.75\columnwidth]{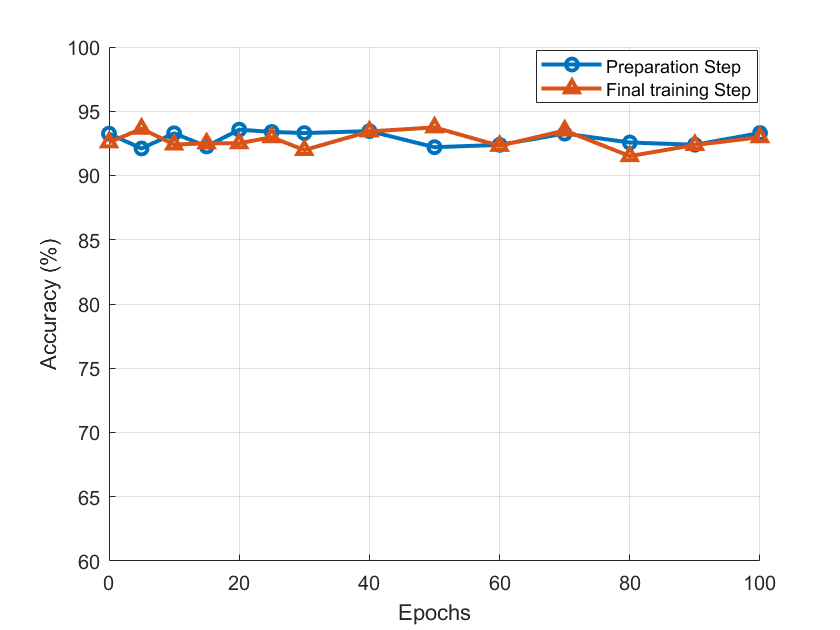}
  \caption{Effect of preparation Step and final training step in PDLRecover}
  \label{fig1}
\vspace{-2.0em}
\end{figure}
Fig. \ref{fig1} illustrates the accuracy variations of the final model across different epochs during the preparation step and the final training step. As shown in Fig. \ref{fig1}, the accuracy curves for both steps are close in most epochs, fluctuating between $90\%$ and $95\%$. While the accuracy during the preparation step shows minor fluctuations, it remains generally sTable~The accuracy during the final training step is slightly more stable, but the difference is minimal. Overall, the model maintains high accuracy throughout the training process, indicating that both the preparation step and the final training step enable the model to effectively learn and sustain high predictive performance.

Furthermore, the number of epochs in both steps has little impact on overall model performance, with excessive epochs merely increasing memory usage, computational resources, and time. Thus, only minimal additional training is required to achieve optimal performance, making it unnecessary to set a large number of epochs to ensure model stability. PDLRecover sets both the preparation step and the final training step to $25$ epochs each, ensuring model performance stability while optimizing resource and time efficiency.

\subsection{Comparisons}
Table~\ref{The accuracy of unlearning methods, comparison} provides a comparative accuracy analysis for various unlearning methods across three datasets: MNIST, FashionMNIST, and HAR. It can be seen that our PDLRecover demonstrates significant improvement in accuracy, particularly on the FashionMNIST and HAR datasets. Specifically, for FashionMNIST, PDLRecover achieves an accuracy of $93.2\%$, which is markedly higher than the accuracies reported by \cite{guo2019certified} ($86.5\%$), \cite{zhang2023fedrecovery} ($87.4\%$), and \cite{cao2023fedrecover} ($85.8\%$), and closely matches the retrain method’s accuracy of $94.9\%$. Similarly, for HAR dataset, PDLRecover achieves an accuracy of $93.4\%$, significantly surpassing the accuracies from \cite{guo2019certified} ($84.4\%$), \cite{zhang2023fedrecovery} ($86.5\%$), and \cite{cao2023fedrecover} ($87.2\%$), and closely matching the retrain method’s accuracy of $94.6\%$. For MNIST dataset, PDLRecover attains an accuracy of $93.9\%$, which is slightly lower than the retrain method ($95.4\%$) and comparable to the accuracies of \cite{guo2019certified} ($95.2\%$) and \cite{zhang2023fedrecovery} ($95.3\%$).

These results indicate that PDLRecover is highly effective in restoring model performance, achieving accuracies that are competitive or superior to existing methods. Moreover, PDLRecover demonstrates robustness against poison attacks, ensuring stable and accurate model recovery across different datasets. This highlights the efficacy and reliability of PDLRecover in practical scenarios, making it a valuable contribution to decentralized learning.

\section{Security Analysis}\label{Security Analysis}
\subsection{Correctness}
\subsubsection*{Theorem}

Let there be $k$ clients, each holding a secret $a_{i,0}$ encoded as the constant term of a degree-$(t_{\text{th}} - 1)$ polynomial $f_i(x) = a_{i,0} + a_{i,1} x + \ldots + a_{i,t_{\text{th}} - 1} x^{t_{\text{th}} - 1}$ defined over a finite field $\mathbb{F}q$. Each client $i$ generates $n$ shares $s{i,1}, \ldots, s_{i,n}$ by evaluating $f_i$ at $n \ge t_{\text{th}}$ mutually distinct nonzero points $x_1, \ldots, x_n \in \mathbb{F}_q$.

We prove that the result of aggregating all clients' secret shares followed by Lagrange interpolation is equal to the sum of the individually reconstructed secrets
\begin{equation}
\text{Lagrange}\left( \sum_i \text{shares}_i \right) = \sum_i \text{Lagrange}(\text{shares}_i).
\end{equation}

\subsubsection*{Step 1: Reconstructing Each Client's Secret}

Each client's secret can be recovered via Lagrange interpolation at $x = 0$
\begin{equation}
a_{i,0} = f_i(0) = \sum_{j=1}^{n} s_{i,j} \cdot \ell_j(0),
\end{equation}
where $\ell_j(0) = \prod_{\substack{1 \le m \le n \ m \ne j}} \frac{-x_m}{x_j - x_m}$ denotes the $j$-th Lagrange basis polynomial evaluated at zero.

\subsubsection*{Step 2: Aggregating Before Reconstruction}

Define the pointwise sum of all clients' polynomials as
\begin{equation}
F(x) = \sum_{i=1}^{k} f_i(x) = \sum_{\ell=0}^{t_{\text{th}} - 1} \left( \sum_{i=1}^{k} a_{i,\ell} \right) x^\ell.
\end{equation}

Evaluating $F$ at $x = 0$ yields the aggregate secret
\begin{equation}
F(0) = \sum_{i=1}^{k} a_{i,0}.
\end{equation}

\subsubsection*{Step 3: Verifying Linearity of Interpolation}

Rewriting the expression for the total reconstructed secret:
\begin{equation}
\sum_{i=1}^{k} a_{i,0} = \sum_{i=1}^{k} \left( \sum_{j=1}^{n} s_{i,j} \cdot \ell_j(0) \right) = \sum_{j=1}^{n} \left( \sum_{i=1}^{k} s_{i,j} \right) \cdot \ell_j(0).
\end{equation}

This matches the result of interpolating the aggregated shares $\sum_{i=1}^{k} s_{i,j}$ evaluated at $x = 0$, that is,
\begin{equation}
F(0) = \sum_{j=1}^{n} \left( \sum_{i=1}^{k} s_{i,j} \right) \cdot \ell_j(0).
\end{equation}

Therefore, Lagrange interpolation is linear with respect to share aggregation.

\subsubsection*{Extension to SS-L-BFGS Approximation}

In PDLRecover, each client $C_j$ computes a local directional gradient approximation using SS-L-BFGS, the extended L-BFGS method, 
\begin{equation}
\hat{g}_t^{(x_j)} = \nabla L^{(x_j)}(\bar{w}_t) + \widetilde{H}^{(x_j)}(\hat{w}_t - \bar{w}_t),
\end{equation}
where $\widetilde{H}^{(x_j)}$ denotes the local approximation HVP, computed using the client's curvature buffer $(\widetilde{W}^{(x_j)}, \widetilde{G}^{(x_j)})$.

Each component of $\hat{g}_t^{(x_j)}$ is secret-shared across clients. The server aggregates these shares and performs Lagrange interpolation to recover the global gradient approximation
\begin{equation}
\hat{g}_t = \sum_{j=1}^{n} \hat{g}_t^{(x_j)} \cdot \ell_j(0) = \sum_{j=1}^{n} \left[ \nabla L^{(x_j)}(\bar{w}_t) + \widetilde{H}^{(x_j)}(\hat{w}_t - \bar{w}_t) \right] \cdot \ell_j(0).
\end{equation}

By linearity of interpolation,
\begin{equation}
\hat{g}_t = \underbrace{\sum_{j=1}^{n} \nabla L^{(x_j)}(\bar{w}_t) \cdot \ell_j(0)}_{\nabla L(\bar{w}_t)} + \underbrace{\sum_{j=1}^{n} \widetilde{H}^{(x_j)}(\hat{w}_t - \bar{w}_t) \cdot \ell_j(0)}_{\widetilde{H} (\hat{w}_t - \bar{w}_t)},
\end{equation}
which produces
\begin{equation}
\hat{g}_t = \nabla L(\bar{w}_t) + \widetilde{H} (\hat{w}_t - \bar{w}t),
\end{equation}
where the global Hessian approximation is defined as $\widetilde{H} = \sum{j=1}^{n} \widetilde{H}^{(x_j)} \cdot \ell_j(0)$.

\subsubsection*{Remarks on Quantization and Field Arithmetic}

To enable arithmetic over a finite field, all real-valued vectors are quantized locally into fixed-point integers using $k$-bit precision and subsequently mapped into the finite field $\mathbb{F}_q$. Both secret sharing and Lagrange interpolation are then performed over $\mathbb{F}_q$ to ensure algebraic consistency.

\subsubsection*{Conclusion}

Both scalar secrets and vector-valued approximations, such as those produced by SS-L-BFGS, preserve the linear homomorphism of shamir secret sharing and the linearity of Lagrange interpolation at $x = 0$. Therefore, we have
\begin{equation}
\text{Lagrange}\left( \sum_j \text{shares}_{j} \right) = \sum_j \text{Lagrange}(\text{shares}_j).
\end{equation}

This confirms that the secure aggregation of local SS-L-BFGS approximations within the PDLRecover framework is mathematically sound and preserves gradient fidelity, without requiring the clients to disclose their individual updates.

\subsection{Privacy Analysis}
In our security analysis, we categorize attackers into two types. The first type is an external attacker. External attackers may attempt to attack buffers stored by clients, which they are not authorized to access. This type of attacks concerns the confidentiality of client content. We will demonstrate that external attackers gain no advantage from such attacks, as they cannot access complete local update information. The second type is an internal attacker, who is authorized to know the subsecrets of other clients. Internal attackers may attempt to attack local updates from honest clients or tamper with the recovered data. Since any client can request data reconstruction, we must ensure that internal attackers cannot access the personal local update information of other clients.

The following theorems prove that our protocol meets the security goals.

\textbf{Theorem 1 (External Attacks).} Suppose an attacker steals the local update subsecret stored by a client. In that case, the attacker cannot obtain complete local update information nor can they reconstruct the local information with any client.

\textit{Proof.} An attacker can target the local update sets stored by individual clients, only a collusion attack involving the compromise of buffers from $k$ clients would allow them to recover the sum of the local updates. However, they cannot recover the individual local updates of any single client. This security feature theoretically ensures the security of local updates.

\textbf{Theorem 2 (Internal Attacks).} Suppose malicious clients attempt to attack local updates from other honest clients or try to tamper with the recovered data. In that case, the malicious clients cannot obtain complete local update information nor they can reconstruct the local information with any client.

\textit{Proof.} Although any client can attack the local update sets stored by other clients, only if more than $k$ malicious clients collude by separately sending the local update subsecret of individual clients can they compute the local updates of honest clients. For the local update secret sharing algorithm in this method, each client saves the local update set $ \{(x_{i}, f_{j}(x_{i}))\} $ during training. However, during the reconstruction of local update information, each client calculates the sum of the subsecrets of the remaining clients $ f(x_{i}) = \sum_{j \in n} f_{j}(x_{i}) $, without revealing the subsecret $ f_{j}(x_{i}) $ shared by an individual client. Therefore, even though a client can reconstruct the polynomial $ f_{j}(x) $ to obtain the personal local update of client $j$, it requires the collusion of more than $k$ malicious clients.


\subsection{Computation and Communication Costs for Clients}
Updating the client computational model incurs both computational and communication costs. These costs can be considered fixed unit expenses, as they do not vary based on the specific round in which the client calculates the model update. In the drop client retraining scenario, the average computational and communication cost is $O(T)$, where $T$ represents the total number of iterations. This is due to the requirement for each client to retrain and update the model during every iteration round.

The number of preparation rounds $T_p$, the periodic round $T_r$, and the number of final training rounds $T_f$ determine the cost of the method we propose. It can be deduced that the average computation and communication cost per client in PDLRecover is $O(T_p + T_f + \left\lceil \frac{T - T_p - T_f}{T_r} \right\rceil)$.

\subsubsection{Bounding the Difference between PDLRecover and Drop Client Retrain in Global Model Recovery}

We outline the assumptions that guide our theoretical evaluation. Next, we display the bound on the difference between the global model that our technique recovered and the drop client retrain model.

\textbf{Assumption 1:} The loss function is $\mu$-strongly convex and L-smooth. Formally, for each client $i$, and for any $\mathbf{w}$ and $\mathbf{w}'$, we have the following inequalities:
\begin{equation}
\langle \mathbf{w} - \mathbf{w}', \nabla L_i(\mathbf{w}) - \nabla L_i(\mathbf{w}') \rangle \geq \mu \|\mathbf{w} - \mathbf{w}'\|^2,    
\end{equation}
\begin{equation}
\langle \mathbf{w} - \mathbf{w}', \nabla L_i(\mathbf{w}) - \nabla L_i(\mathbf{w}') \rangle \geq \frac{1}{\mathcal{L}} \|\nabla L_i(\mathbf{w}) - \nabla L_i(\mathbf{w}')\|^2,    
\end{equation}
where $L_i$ is the loss function for client $i$, $\langle \cdot , \cdot \rangle$ denotes the inner product of two vectors, and $\|\cdot\|$ represents the $\ell_2$ norm of a vector.

\textbf{Assumption 2:} The approximation error of the Hessian-vector product in SS-L-BFGS algorithm is bounded. Formally, each approximate Hessian-vector product satisfies the following condition:
\begin{equation}
\forall i, \forall t, \|\mathbf{\tilde{H}}_t^i (\mathbf{\hat{w}}_t - \mathbf{\bar{w}}_t) + \nabla L_i(\mathbf{\hat{w}}) - \nabla L_i(\mathbf{\bar{w}})\| \leq Z,    
\end{equation}
where $Z$ is a finite positive value.

\textbf{Theorem 1:} Assume the following two conditions are met: all malicious clients have been identified, FedAvg is utilized as an aggregation rule, and the learning rate $\gamma$ fulfills $\gamma \leq \min \left(\frac{1}{\mu}, \frac{1}{\mathcal{L}}\right)$. The global model recovered by our method and the global model obtained by deleting clients for retraining can therefore be distinguished at any iteration $t > 0$ as follows:
\begin{equation}
\|\mathbf{\hat{w}}_t - \mathbf{\bar{w}}_t\| \leq (\sqrt{1 - \gamma \mu})^t \|\mathbf{\hat{w}}_0 - \mathbf{\bar{w}}_0\| + \frac{1 - (\sqrt{1 - \gamma \mu})^t}{1 - \sqrt{1 - \gamma \mu}} \gamma M,
\end{equation}
where $\mathbf{\hat{w}}_t$ and $\mathbf{\bar{w}}_t$ are the global models recovered by PDLRecover and drop client retrain, respectively, in iteration $t$.

\textbf{Proof:} PDLRecover is to recursively bound the difference in each iteration.

According to Theorem 1, we have
\begin{equation}
\lim_{t \to \infty} \|\mathbf{\hat{w}}_t - \mathbf{\bar{w}}_t\| \leq \frac{\gamma Z}{1 - \sqrt{1 - \gamma \mu}}.
\end{equation}

Additionally, we derive the following corollary:

\textbf{Corollary 1:} When the SS-L-BFGS algorithm can accurately compute the Hessian-vector product , the bound on the difference between the global model recovered by PDLRecover and the one recovered by drop client retrain is given by
\begin{equation}
\|\mathbf{\hat{w}}_t - \mathbf{\bar{w}}_t\| \leq (\sqrt{1 - \gamma \mu})^t \|\mathbf{\hat{w}}_0 - \mathbf{\bar{w}}_0\|
.
\end{equation}

Therefore, the global model recovered by PDLRecover converges to the one recovered by drop client retrain, i.e., $\lim_{t \to \infty} \mathbf{\hat{w}}_t = \lim_{t \to \infty} \mathbf{\bar{w}}_t$.

\subsubsection{Trade-off between Difference Bound and Computation/Communication Costs}

Based on Corollary 1, we have
\begin{equation}
\|\mathbf{\hat{w}}_T - \mathbf{w}_T\| \leq (\sqrt{1 - \gamma \mu})^T \|\mathbf{\hat{w}}_0 - \mathbf{w}_0\|
,    
\end{equation} 
when PDLRecover runs for $T$ rounds. As $T$ increases, the difference bound decreases exponentially. The computation and communication costs of PDLRecover are linear with $T$. Therefore, as costs increase, the difference bound decreases exponentially. In other words, we observe an accuracy-cost trade-off for PDLRecover: The recovered global model becomes more accurate (i.e., closer to the drop client retrain model) if more computational and communication cost are spent, that is, the computation and communication costs for clients increase accordingly.

\section{Conclusion}\label{conclustion}
In this paper, we proposed PDLRecover, a novel decentralized unlearning framework that enables secure and efficient recovery from poison attacks without requiring clients to access or reveal each other’s model updates. PDLRecover reconstructs the global model via Lagrange interpolation while ensuring complete local privacy. Each client independently computes its contribution to the recovery direction using only private gradient and curvature information, and the global model update is collaboratively reconstructed without exposing any individual values. Our theoretical analysis and empirical results demonstrate that PDLRecover not only preserves the confidentiality of historical updates but also enables accurate reconstruction of the global model in the presence of malicious or dropped clients. Unlike traditional detection-based defenses, PDLRecover provides a proactive, privacy-preserving solution to mitigating poison attacks in decentralized environments.

In future work, we will extend PDLRecover with advanced secure aggregation techniques to better address a wider variety of attack types and operating environments. We will also focus on developing multi-layered defense mechanisms to further bolster the security and recoverability of the global model. Additionally, we aim to investigate the potential for model recovery with reduced dependency on historical data, which could lower storage requirements and enhance the efficiency of model updates.

\bibliographystyle{IEEEtran}
\bibliography{reference}
\vspace{12pt}

\end{document}

%% file: Table_Alg/Notation.tex
\begin{table}[t]
\centering
\renewcommand\arraystretch{1.4}
\caption{Notations}
\begin{tabular}{|c|l|}
\hline
\textbf{Notation} & \textbf{Description} \\ \hline
$n$ & Number of clients\\ \hline
$m$ & Number of malicious clients \\ \hline
$t$ & Iteration index \\ \hline
$i$ & Client index \\ \hline
$\gamma$ & Learning rate \\ \hline
$T$ & Total number of rounds \\ \hline
$T_p$ & Number of preparation step \\ \hline
$T_r$ & Index of periodic step \\ \hline
$T_f$ & Number of final exact training \\ \hline
$s$ & Buffer size of the L-BFGS algorithm \\ \hline
$\mathbf{\bar{w}}_t$ & Original global model at iteration $t$ \\ \hline
$\mathbf{\hat{w}}_t$ & Recovery global model at iteration $t$ \\ \hline
$\nabla L_{i}(\mathbf{\bar{w}}_t)$ & Original local update for client $i$ at iteration $t$ \\ \hline
$\nabla L_{i}(\mathbf{\hat{w}}_t)$ & Recovery local update for client $i$ at iteration $t$ \\ \hline
$\nabla L(\mathbf{\bar{w}}_t)$ & Sum of original local update for client $i$ at iteration $t$ \\ \hline
$\nabla L(\mathbf{\hat{w}}_t)$ & Sum of recovery local update at iteration $t$ \\ \hline
$\nabla L^{(x_i)}(\hat{w}_t)$  & Sub-secret recovery local update for client $i$ at iteration $t$ \\ \hline
$\widetilde{\mathbf{H}_t}$ & Estimated Hessian matrix at iteration $t$ \\ \hline
$\widetilde{G}^{(x_i)} $& Local update difference buffer for client $i$\\\hline
$\widetilde{W}$&Global model difference buffer\\\hline
\end{tabular}
\label{Notation}
\end{table}

%% file: Table_Alg/PDLRecover.tex
\begin{algorithm}[!t] 
\caption{PDLRecover}
\label{PDLRecover}
\begin{algorithmic}[1]
\Require 
Clients $C_r = { C_i \mid m + 1 \le i \le n }$; initial model $\bar{w}_0$; learning rate $\gamma$; original global models ${ \mathbf{\hat{w}}_0, \mathbf{\hat{w}}_1, \ldots, \mathbf{\hat{w}}_T }$; sub-secret of client $j$'s local model ${ \nabla L^{(x_i)}(\bar{w}_0), \nabla L^{(x_i)}(\bar{w}_1), \ldots, \nabla L^{(x_i)}(\bar{w}_t) }$; periodic step interval $T_r$; final exact steps $T_f$; SS-L-BFGS buffer size $s$; share points ${ x_i }$; total rounds $T$.
\Ensure Final recovered model $\hat{w}_T$

\State $\hat{w}_0 \gets \bar{w}_0$

\For{$t = 0$ to $T_p - 1$}
    \State $\hat{w}_{t+1} \gets$ \Call{ExactUpdate}{$C_r, \hat{w}_t, \gamma$} \Comment{preparation step}
\EndFor

\For{$t = T_p$ to $T - T_f - 1$}
    \If{$(t - T_p + 1) \bmod T_r = 0$} \Comment{Periodic step}
        \State  $\nabla L^{(x_i)}(\hat{w}_t)$, $\hat{w}_{t+1} \gets$ \Call{ExactUpdate}{$C_r, \hat{w}_t, \gamma$} 

        \For{each client $C_i$}
            \State $\Delta G^{(x_i)}_t = \nabla L^{(x_i)}(\hat{w}_t) - \nabla L^{(x_i)}(\bar{w}_t)$
            \State $\Delta W_t = \hat{w}_t - \bar{w}_t$
            \State $\widetilde{G}^{(x_i)}_t \gets \widetilde{G}^{(x_i)}_t \cup \{ \Delta G^{(x_i)}_t \}$
            \State $\widetilde{W}_t \gets \widetilde{W}_t\cup \{ \Delta W_t \}$
        \EndFor
    \Else 
        \For{each client $C_i$}
            \State $\widetilde{H}_i^t \mathbf{v} \gets$ \Call{SS-L-BFGS}{$\widetilde{W}_t, \widetilde{G}^{(x_i)}_t, \hat{w}_t - \bar{w}_t$}
            \State $\hat{g}^{(x_i)}_t = \nabla L^{(x_j)}(\bar{w}_t) + \widetilde{H}_i^t\mathbf{v}$
        \EndFor
        \State $\hat{g}^t = \sum_{j=1}^{t} \hat{g}_t^{(x_i)} \cdot \prod_{\substack{1 \leq i \leq t \\i \ne j}} \frac{0 - x_i}{x_j - x_i}$

        \State $\hat{w}_{t+1} \gets \hat{w}_t - \frac{\gamma}{n} \cdot \hat{g}^t$
    \EndIf
\EndFor

\For{$t = T - T_f$ to $T - 1$} \Comment{Final exact recovery}
    \State $\hat{w}_{t+1} \gets$ \Call{ExactUpdate}{$C_r, \hat{w}_t, \gamma$}
\EndFor

\State \Return $\hat{w}_T$

\vspace{1em}
\State \textbf{ExactUpdate:}
\State \textbf{Client $i$ computes and share local updates:}
\State \indent $g_i = \nabla L_i(\mathbf{w}) = \frac{1}{n} \sum_{i=1}^{n} \nabla L_{ij}(\mathbf{w})$
\State \indent Generate polynomial for local updates $\nabla L(\mathbf{\bar{w}}_{t}^{i})$
\State \indent Compute secret shares $ (x_{i}^{t}, f_{j}^{t}(x_{i})) $ for each client $i$

\State \textbf{Client $i$ aggregates received shares:}
\State \indent Upon receiving shares $ \{(x_{i}, f_{j}^{t}(x_{i}))\} $ from all clients,
\State \indent Compute polynomial sum $ \nabla L^{(x_j)}(\bar{w}_t) = \sum_{j \in n} f_{j}^{t}(x_{i}) $

\State \textbf{Client $i$ reconstructs local model update:}
\State \indent $\nabla L(\mathbf{\hat{w}}_{t}) = \sum_{i=1}^{t} l_{i}^{t} \prod_{1 \leq j \leq t, j \neq i} \frac{x - j}{i - j}$

\State \textbf{Client $i$ updates model:}
\State \indent $\mathbf{\hat{w}}_{t+1} = \mathbf{\hat{w}}_{t} - \frac{\gamma \cdot \nabla L(\mathbf{\hat{w}}_{t})}{n}$
\end{algorithmic}
\end{algorithm}

%% file: Table_Alg/L-BFGS.tex
\begin{algorithm}[t] 
\caption{SS-L-BFGS}
\label{alg:lbfgs_share}
\begin{algorithmic}[1] 
\Require 
Local model difference buffer $\widetilde{W}^{(x_j)} = [\Delta \mathbf{w}^{(x_j)}_1, \cdots, \Delta \mathbf{w}^{(x_j)}_s]$, Local gradient difference buffer $\widetilde{G}^{(x_j)} = [\Delta \mathbf{g}^{(x_j)}_1, \cdots, \Delta \mathbf{g}^{(x_j)}_s]$, A direction vector $\mathbf{v}$
\Ensure Local approximated HVP $\widetilde{\mathbf{H}}^{(x_j)} \mathbf{v}$
\State $\mathbf{A} = (\widetilde{W}^{(x_j)})^T \widetilde{G}^{(x_j)}$
\State $\mathbf{D} = \text{diag}(\mathbf{A})$ 
\State $\mathbf{M} = \text{lower-triangular}(\mathbf{A})$ 
\State $\rho = (\Delta \mathbf{g}^{(x_j)}_{s})^T \Delta \mathbf{w}^{(x_j)}_{s} /(\Delta \mathbf{w}^{(x_j)}_{s})^T \Delta \mathbf{w}^{(x_j)}_{s}$
\State $\mathbf{p} = 
\begin{bmatrix} 
    -\mathbf{D} & \mathbf{M}^T \\
    \mathbf{M} & \rho (\widetilde{W}^{(x_j)})^T \widetilde{W}^{(x_j)}
\end{bmatrix}^{-1}
\begin{bmatrix} 
    \widetilde{G}^{(x_j)} \mathbf{v} \\
    \rho (\widetilde{W}^{(x_j)})^T \mathbf{v} 
\end{bmatrix}
$
\State $\widetilde{\mathbf{H}}^{(x_j)} \mathbf{v} = \rho \mathbf{v} - \begin{bmatrix} \widetilde{G}^{(x_j)} & \rho \widetilde{W}^{(x_j)} \end{bmatrix} \cdot \mathbf{p}$
\State \Return $\widetilde{\mathbf{H}}^{(x_j)} \mathbf{v}$
\end{algorithmic}
\end{algorithm}